\documentclass[paper]{JHEP}
\usepackage[centertags]{amsmath}
\usepackage{amsfonts} \usepackage{amssymb} \usepackage{amsthm}
\usepackage{graphicx}
\usepackage{psfrag}

\newcommand{\ket}[1]{|{#1}\rangle}
\def\one{{\rm 1\kern -.9mm l}}

\def\beq{\begin{equation}}
\def\eeq{\end{equation}}
\def\beqa{\begin{eqnarray}}
\def\eeqa{\end{eqnarray}}

\newcommand{\comm}[2]{\left[#1,#2\right]}
\newcommand{\Tr}{\mathrm{Tr}\,}
\def\tr{{\rm tr}\,}

\newcommand{\lvev}{\langle\hskip -6pt\langle\hskip 4pt}
\newcommand{\rvev}{\hskip 4pt\rangle\hskip -6pt\rangle}
\newcommand{\bea}{\begin{eqnarray}}
\newcommand{\ena}{\end{eqnarray}}

\usepackage{graphicx}
\usepackage{dcolumn}
\def\ii{\mathrm{i}}
\def\ee{\mathrm{e}}
\title{$\mathbf{\mathcal{N}=1/2}$ gauge theory and its instanton moduli space from
open strings in  R-R background\thanks{Work partially supported by the European
Commission's Improving Human Potential program under the contract HPRN-CT-2000-00131,
``The quantum structure of spacetime and the geometric nature of fundamental interactions'',
and by the Italian M.I.U.R under the contract P.R.I.N. 2003023852,
``Physics of fundamental interactions: gauge theories, gravity and strings''.}}

\author{
Marco Bill\'o
,
Marialuisa Frau
,
Igor Pesando
\\
Dipartimento di Fisica Teorica, Universit\`a di Torino\\
and Istituto Nazionale di Fisica Nucleare - sezione di Torino \\
via P. Giuria 1, I-10125 Torino, Italy
}
\author{
Alberto Lerda%
\thanks{E-mails: \tt{billo,frau,ipesando,lerda@to.infn.it}}\\
Dipartimento di Scienze e Tecnologie Avanzate\\
Universit\`a del Piemonte Orientale, I-15100 Alessandria, Italy\\
and Istituto Nazionale di Fisica Nucleare  - sezione di Torino \\
via P. Giuria 1, I-10125 Torino, Italy
}

\abstract{We derive the four dimensional $\mathcal{N}=1/2$ super Yang-Mills theory
from tree-level computations in RNS open string theory with insertions of
closed string Ramond-Ramond vertices.
We also study instanton configurations in this gauge theory and their ADHM moduli
space, using systems of D3 and D(--1) branes in a R-R background.}

\preprint{DFTT-7/2004}

\begin{document}

\section{Introduction}
\label{sec:intro}
The study of the effects of non trivial closed string backgrounds
on the low-energy dynamics of open strings and D-branes has
attracted a lot of interest in the last few years for many
reasons. Among the simplest, yet non-trivial, possibilities that
have been considered are the backgrounds in which some of the
antisymmetric tensors of the closed string spectrum acquire
a constant non-zero value. For example a constant profile for the
$B_{\mu\nu}$ field of the NS-NS sector modifies the open string dynamics
by introducing new couplings and interactions
which can also be interpreted in terms of a non-commutative
deformation of the space where the strings propagate
\cite{Seiberg:1999vs}. Field theories, and in particular gauge theories,
defined on non-commutative spaces were the subject of
vast investigations even before the relation with string theory was
realized, but it was only after the connection with
the propagation of strings in a $B_{\mu\nu}$ background was
exhibited that many properties of non-commutative theories
were elucidated and put in a broader perspective.

More recently, other kinds of closed string backgrounds have been considered.
In particular, in the context of Type II B string theory compactified
on a Calabi-Yau threefold, the effects of the presence of a constant non-vanishing
graviphoton field strength $C_{\mu\nu}$ have been analyzed by several
authors \cite{deBoer:2003dn,Ooguri:2003qp,Ooguri:2003tt,Seiberg:2003yz,Berkovits:2003kj}.
A graviphoton background can be obtained by wrapping
the 5-form field strength of the R-R sector of Type II B string theory on a 3-cycle of the
internal Calabi-Yau manifold, and a consistent possibility in Euclidean space
is to take a $C_{\mu\nu}$
with a definite duality, for example anti self-dual.
A constant anti self-dual graviphoton
field strength induces a deformation of the four dimensional superspace
in which the fermionic coordinates are no longer
anticommuting Grassmann variables but become elements of a Clifford
algebra \cite{Ferrara:2000mm,Klemm:2001yu,deBoer:2003dn,Ooguri:2003qp,
Ooguri:2003tt,Seiberg:2003yz,Berkovits:2003kj,Terashima:2003ri}
\begin{equation}
\big\{\theta^{\alpha},\theta^{\beta}\big\} \,=\,
\big\{\theta^{\alpha},\bar\theta^{\dot\beta}\big\} \, = \, 0 ~~,~~
\big\{\bar\theta^{\dot\alpha},\bar\theta^{\dot\beta}\big\} = C^{\dot\alpha\dot\beta}
\label{thetatheta}
\end{equation}
where
$C^{\dot\alpha\dot\beta}=\frac{1}{4}C_{\mu\nu}(\bar\sigma^{\mu\nu})^{\dot\alpha\dot\beta}$.
The non-vanishing anticommutator in (\ref{thetatheta}) breaks the four dimensional Lorentz
group $\mathrm{SU}(2)_{\mathrm{L}}\times \mathrm{SU}(2)_{\mathrm{R}}$
to $\mathrm{SU}(2)_{\mathrm{L}}$, and reduces the number of preserved supercharges
by a factor of two. Therefore, a graviphoton background deforms
a $\mathcal{N}=1$ field theory in four dimensions
to a $\mathcal{N}={1}/{2}$ theory with only two preserved
supercharges and new types of interactions that are induced by the non-anticommutative
structure of the superspace.
Supersymmetric field theories based on non-anticommutative
superspaces and their renormalization properties
have been largely studied in the recent past from different points of view
\cite{Britto:2003aj,Grisaru:2003fd,Araki:2003se,Britto:2003kg,Lunin:2003bm,
Berenstein:2003sr,Alishahiha:2003kg}.
More recently, also the instanton configurations of the $\mathcal{N}={1}/{2}$
gauge theory have been analyzed \cite{Imaanpur:2003jj,Grassi:2003qk,Britto:2003uv}
and generalizations with extended supersymmetry have been
proposed \cite{Ferrara:2003xk,Ivanov:2003te}.

Even if the non-anticommutative algebra (\ref{thetatheta}) has
a direct string theory interpretation as we mentioned above,
so far most of the analysis of the $\mathcal{N}={1}/{2}$
field theories has been carried out by exploiting the superspace deformations
that are induced by the graviphoton, without making explicit reference
to string theory. In this paper we fill this gap and show that the
$\mathcal{N}={1}/{2}$ gauge theories in four dimensions can be
also obtained directly from string theory by computing, in the standard RNS
formalism, scattering amplitudes in the presence of a R-R background with constant
field strength.
It is a common belief that the RNS formalism is not suited to
deal with a R-R background; while this is true in general, it is not exactly
so when the R-R field strength is constant. In fact, in this case one can
represent the background by a R-R vertex operator at {\it zero}
momentum which in principle can be repeatedly inserted inside disk correlation
functions among open string vertices without affecting
their dynamics. As we will see explicitly in section \ref{sec:D3},
the integrals on the world-sheet variables that arise from these insertions turn out to
be elementary and thus the effects of the R-R background on the open string dynamics
can be explicitly computed in this way. Even though this method is intrinsically
perturbative, in the field theory limit $\alpha'\to 0$ the procedure
stops after the first step and so the results one obtains in this way are
exact in this limit. This is a consequence of the fact
that the R-R graviphoton background modifies the fermionic sector
of the superspace as shown in (\ref{thetatheta}) and induces
a star product which, when expanded,
contains only a finite number of background insertions as a consequence of the
fermionic nature of the $\bar\theta$'s coordinates.

This paper is organized as follows: in section \ref{sec:D3} we
briefly review how to engineer the four dimensional $\mathcal{N}=1$ super
Yang-Mills theory with gauge group $\mathrm{U}(N)$ in terms of
$N$ fractional D3 branes in the orbifold
$\mathbb{R}^6/(\mathbb{Z}_2\times\mathbb{Z}_2)$ and, by explicitly computing
tree-level scattering amplitudes of open strings on
disks with the insertion of a R-R vertex operator, we discuss
the deformations induced by a graviphoton background
on the world-volume theory. In particular, taking the field theory limit
$\alpha'\to 0$ we can recover the action of the $\mathcal{N}={1}/{2}$ super
Yang-Mills theory directly from string computations.
In section \ref{sec:D-1} we extend this analysis to a system of
$N$ D3 and $k$ D(--1) branes in order to describe the $k$
instanton sector of this gauge theory and, generalizing
our previous results \cite{Billo:2002hm}, we discuss how the structure of
the instanton moduli space and the ADHM constraints are modified
by the R-R background. This analysis involves the explicit
calculation of open string amplitudes on disks which have at least
a part of their boundary on the D-instantons and which may also contain insertions of
the graviphoton vertices. In section \ref{sec:def_inst}
we show that these mixed disks are the sources for the
(super)-instantons of the $\mathcal{N}={1}/{2}$ $\mathrm{U}(N)$ gauge theory.
In particular, we compute the emission amplitude of the gluon field from a mixed disk
in presence of a R-R vertex operator. {F}rom this amplitude, in analogy with what happens in the
closed string with the boundary state \cite{DiVecchia:1997pr}, we deduce the
leading term in the large distance expansion of the gluon profile in the
singular gauge and find how the graviphoton background affects the
instanton solution, confirming in this way the general structure that has been recently
uncovered in the regular gauge
\cite{Imaanpur:2003jj,Grassi:2003qk,Britto:2003uv}.
Finally, in the appendix we list our conventions and collect
technical details and useful formulas for our calculations.

\vspace{0.5cm}
\section{The $\mathcal{N}=1/2$ gauge theory from open strings in a R-R background}
\label{sec:D3}
In this section we show how the gauge theory
deformations induced by a graviphoton background can be derived
directly from string theory. Let us begin by considering the pure
${\mathcal N}=1$ SYM theory in four (euclidean) dimensions with
gauge group $\mathrm{U}(N)$ whose action is given by%
\footnote{For our conventions see appendix \ref{app:sub:target}.}
\begin{equation}
S= \frac{1}{g^2_{\rm YM}} \int d^4x \,{\rm
Tr}\Big(\frac{1}{2}F_{\mu\nu}^2 -2\bar\Lambda_{\dot\alpha}\bar
D\!\!\!\!/^{\,\dot\alpha \beta} \Lambda_\beta\Big)~~.
\label{n1}
\end{equation}
As is well-known this action describes the low-energy dynamics on
a stack of $N$ (fractional) D3 branes placed at the singularity of
the orbifold $\mathbb{R}^6/(\mathbb{Z}_2\times\mathbb{Z}_2)$,
whose massless excitations are the gauge boson $A_\mu$ and the
gauginos $\Lambda^{\alpha}$ and $\bar\Lambda_{\dot\alpha}$. These
are represented by the following open string vertex operators
\begin{equation}
V_A(y;p) = (2\pi\alpha')^{\frac{1}{2}}\,\frac{A_\mu(p)}{\sqrt
2}\,\psi^\mu(y)\,{\rm e}^{-\phi(y)}\,{\rm e}^{i \sqrt{2\pi\alpha'}p\cdot X(y)}~~,
\label{amu}
\end{equation}
\begin{equation}
V_\Lambda(y;p)=
(2\pi\alpha')^{\frac{3}{4}}\,\Lambda^{\alpha}(p)\,S_\alpha(y)\,S^{(-)}(y)\,
{\rm e}^{-\frac{1}{2}\phi(y)}\,{\rm e}^{i \sqrt{2\pi\alpha'}p\cdot
X(y)}~~,
\label{lambda}
\end{equation}
and
\begin{equation}
V_{\bar\Lambda}(y;p) =
(2\pi\alpha')^{\frac{3}{4}}\,\bar\Lambda_{\dot\alpha}(p)\,S^{\dot\alpha}(y)\,S^{(+)}(y)\,
{\rm e}^{-\frac{1}{2}\phi(y)}\,{\rm e}^{i \sqrt{2\pi\alpha'}p\cdot
X(y)}~~,
\label{barlambda}
\end{equation}
with $p^2=0$. In these vertices $\phi$ is the (chiral) boson of
the superghost bosonization formulae \cite{Friedan:1985ge}, $X^\mu$ and
$\psi^\mu$ are the bosonic and fermionic string coordinates along
the longitudinal directions of the D3 branes, $S_\alpha\,S^{(-)}$
and $S^{\dot\alpha}\,S^{(+)}$ are the spin field components which
survive the GSO and orbifold projections (see appendix \ref{app:sub:correlators},
in particular eq. (\ref{orbi_spin_fields})),
and $y$ is a point on
the real axis. Finally, the factors of $(2\pi\alpha')$ in
(\ref{amu}) -- (\ref{barlambda}) have been introduced to assign
canonical dimensions to the polarizations, namely (length)$^{-1}$
to the gauge boson and (length)$^{-\frac{3}{2}}$ to the gauginos,
keeping, as customary, the vertex operators dimensionless.%
\footnote{Notice that the polarization $A_\mu(p)$ has the {\sl same} dimension of
the the field $A_\mu(x)$ because the Fourier transform is taken
w.r.t. to the adimensional momentum $k=\sqrt{2\pi\alpha'}p$.}
Note that the above polarizations include also
$\mathrm{U}(N)$ Chan-Paton factors $T^I$ in the adjoint representation, which we
normalize as
\begin{equation}
{\rm Tr}\big(T^I\,T^J\big)=\frac{1}{2}\,\delta^{IJ}~~.
\label{normalization}
\end{equation}

The various interaction terms in the super Yang-Mills action
(\ref{n1}) can be obtained by
computing the field theory limit $\alpha'\to 0$ of string
scattering amplitudes among the vertex operators
(\ref{amu}) -- (\ref{barlambda}). For example, the (color ordered)
amplitude among one gauge boson and two gauginos is
\begin{equation}
\lvev V_{\bar\Lambda}\, V_{A}\,V_{\Lambda}\rvev \,\equiv\,
C_4\,\int\frac{\prod_{i} dy_i}{dV_{\rm CKG}}~ \big\langle
V_{\bar\Lambda}(y_1;p_1)\, V_{A}(y_2;p_2)\,V_{\Lambda}(y_3;p_3) \big\rangle~~,
\label{ampl33}
\end{equation}
where $dV_{\rm CKG}$ is the $\mathrm{SL}(2,\mathbb{R})$ invariant
volume element and $C_4$ is the topological normalization of a disk with the
boundary conditions of a D3 brane given by~\cite{DiVecchia:1996uq,Billo:2002hm}
\begin{equation}
C_4 = \frac{1}{\pi^2{\alpha'}^2}\,\frac{1}{g_{\rm YM}^2}~~.
\label{C4}
\end{equation}
Using the contraction formulas of appendix \ref {app:sub:correlators} and fixing the
positions of the vertices to three arbitrary points so that
\begin{equation}
dV_{\rm CGK}= \frac{dy_a\,dy_b\,dy_c}{(y_a-y_b)(y_b-y_c)(y_c-y_a)}~~,
\label{projvolume}
\end{equation}
it is easy to find that
\begin{equation}
\lvev V_{\bar\Lambda}\, V_{A}\,V_{\Lambda}\rvev =
-\,\frac{2\,\ii}{g_{\rm YM}^2}\, {\rm Tr}\left(\bar
\Lambda_{\dot\alpha}(p_1)\,{\bar A\!\!\!/}^{\dot\alpha \beta}(p_2)
\,\Lambda_{\beta}(p_3)\right)
\label{ampl331}
\end{equation}
where we have understood the $\delta$-function of momentum
conservation (we will do the same also in the following). Note
that all factors of $\alpha'$ from the disk normalization $C_4$
and the vertices cancel out, so that this result survives in the
field theory limit. The complete coupling among a gauge boson and
two gauginos is obtained by adding to (\ref{ampl331}) the other
inequivalent color order of the fields and thus the term $\mathrm{Tr}\big(
\bar\Lambda_{\dot\alpha}\big[{\bar A\!\!\!/}^{\dot\alpha \beta},
\Lambda_{\beta}\big]\big)$ of the action (\ref{n1}) is recovered.
Proceeding systematically in this way, one can check that indeed all
interaction terms in (\ref{n1}) arise from the $\alpha'\to 0$ limit of
scattering amplitudes%
\footnote{Remember that in Euclidean
space the 1PI part of a scattering amplitude is equal to {\it
minus} the corresponding interaction term in the action.}
among the vertices (\ref{amu}) -- (\ref{barlambda}).

It is interesting to note that the quartic interactions in
${\rm Tr}F_{\mu\nu}^2$ can be decoupled by introducing an
auxiliary antisymmetric tensor $H_{\mu\nu}$ of definite duality
(say, anti self-dual), in the adjoint representation and with
dimension (length)$^{-2}$, which we can write as
\begin{equation}
H_{\mu\nu} = H_c\,\bar\eta^c_{\mu\nu} \label{hmunu}
\end{equation}
where $\bar\eta^c_{\mu\nu}$ are the anti self-dual 't Hooft
symbols%
\footnote{This choice of duality is related to the fact that
later we will introduce an anti self-dual graviphoton background.}.
In fact the action (\ref{n1}) is equivalent to the following one
\begin{equation}
\label{n11}
\begin{aligned}
S'&= \frac{1}{g^2_{\rm YM}} \int d^4x ~\mathrm{Tr}
\Big\{\big(\partial_\mu A_\nu-\partial_\nu A_\mu\big)
\partial^\mu A^\nu \,+\, 2{\rm i}\, \partial_\mu A_\nu
\big[A^\mu,A^\nu\big] \\
&~~~~~~~~~~~~~-\,2\bar\Lambda_{\dot\alpha}\bar
D\!\!\!\!/^{\,\dot\alpha \beta} \Lambda_\beta +H_cH^c +
H_c\,\bar\eta^c_{\mu\nu}\,\big[A^\mu,A^\nu\big] \Big\}~~,
\end{aligned}
\end{equation}
which contains only cubic interaction terms. As shown in
\cite{Billo:2002hm} for the analogous case of D-instantons, also the
auxiliary field $H_{\mu\nu}$ of the D3 branes admits a
representation in string theory since it can be effectively
associated to the following vertex operator (in the 0 superghost
picture)
\begin{equation}
V_H(y;p) =
(2\pi\alpha')\,\frac{H_{\mu\nu}(p)}{2}\,\psi^\nu\psi^\mu(y)\,{\rm
e}^{i \sqrt{2\pi\alpha'} p \cdot X(y)}~~,
\label{vhmunu}
\end{equation}
which has conformal weight 1 if $p^2=0$. The factor of
$(2\pi\alpha')$ has been introduced in order to assign the
required dimension to the polarization $H_{\mu\nu}$, which
includes also the appropriate $\mathrm{U}(N)$ Chan-Paton factor.

It is very easy to verify that all terms in the action $S'$ can be
obtained from the limit $\alpha'\to 0$ of string amplitudes. For
example the (color ordered) coupling among the auxiliary field $H$
and two gauge bosons is given by
\begin{equation}
\frac{1}{2}\,\lvev V_{H}\,V_{A}\,V_{A} \rvev = -\frac{1}{g_{\rm YM}^2}\,
\mathrm{Tr}\,\Big(H_{\mu\nu}(p_1)A^\mu(p_2)A^\nu(p_3)\Big)
\label{amplHAA}
\end{equation}
where the symmetry factor of $\frac{1}{2}$ has been introduced to
account for the presence of two alike fields. Again all factors of
$\alpha'$ cancel out and this result survives in the field theory
limit. Adding to (\ref{amplHAA}) the amplitude with the other
inequivalent color order of the three vertex operators, one
reconstructs the last term of (\ref{n11}). Furthermore, one can
easily check that all other amplitudes involving $V_H$ vanish in
the limit $\alpha'\to 0$, so that the complete field theory result
is given by the action (\ref{n11}).

\vspace{0.2cm}
\subsection{The effects of the graviphoton background}
\label{subsec:RR_D3}
We now analyze the deformations of this $\mathcal{N}=1$ gauge
theory that are induced by a graviphoton background with constant
field strength. This background is usually described by a constant
antisymmetric tensor $C_{\mu\nu}$ with definite duality (here we
take it to be anti self-dual) which is responsible for a
non-anticommutative deformation of the $\mathcal{N}=1$ superspace
\cite{deBoer:2003dn,Ooguri:2003qp,Ooguri:2003tt,Seiberg:2003yz,Berkovits:2003kj}.
{F}rom the string point of view $C_{\mu\nu}$ corresponds to
a R-R field strength; more precisely it is the R-R 5-form $F^{(5)}$
of type II B string theory, wrapped around the 3-cycle of the
internal Calabi-Yau space. In our case the internal space is the
orbifold $\mathbb{R}^6/(\mathbb{Z}_2 \times \mathbb{Z}_2)$ and the
constant graviphoton field strength is described by the following
closed string vertex operator (in the $(-1/2,-1/2)$ superghost picture)
\begin{equation}
\label{RRvop}
V_{\mathcal{F}}(z,\bar z) =
\mathcal{F}_{\dot\alpha\dot\beta}\, S^{\dot\alpha}(z)
S^{(+)}(z) \ee^{-\frac{1}{2}\phi(z)}\,\, {\widetilde
S}^{\dot\beta}(\bar z) {\widetilde{S}}^{(+)}(\bar z)
\ee^{-\frac{1}{2}\tilde\phi(\bar z)}
\end{equation}
where the dimensionless polarization is a symmetric bi-spinor
\begin{equation}
\mathcal{F}_{\dot\alpha \dot\beta} = \mathcal{F}_{\dot\beta
\dot\alpha}~~.
\label{fRR}
\end{equation}
In the vertex (\ref{RRvop}) the tilde
denotes the right movers, and $z$ a point in the upper-half complex
plane. As we will see later, the tensor $C_{\mu\nu}$ that is
usually considered in the literature turns out to be proportional
to $\mathcal{F}_{\dot\alpha \dot\beta}
\left(\bar\sigma_{\mu\nu}\right)^{\dot\alpha \dot\beta}$, which is
clearly anti self-dual. Notice that the vertex operator
(\ref{RRvop}) does not have a $\ee^{\ii \sqrt{2\pi\alpha'} p \cdot X}$ term. In
fact, we are considering a {\it constant} background and hence
$p=0$. For this reason, as we shall explicitly see in the following,
it is possible to use the RNS formulation
of string theory and compute the effects of this R-R background on
the gauge theory by evaluating scattering amplitudes on disks with
insertions of the vertex operator (\ref{RRvop}) in the interior.

\FIGURE{
\begin{minipage}[b]{0.45\linewidth}
\centering
{
\psfrag{(a)}{\emph{(a)}}
\psfrag{L}{$\Lambda$}
\psfrag{A}{$A_\mu$}
\psfrag{FRR}{$\mathcal{F}$}
\includegraphics[width=0.5\textwidth]{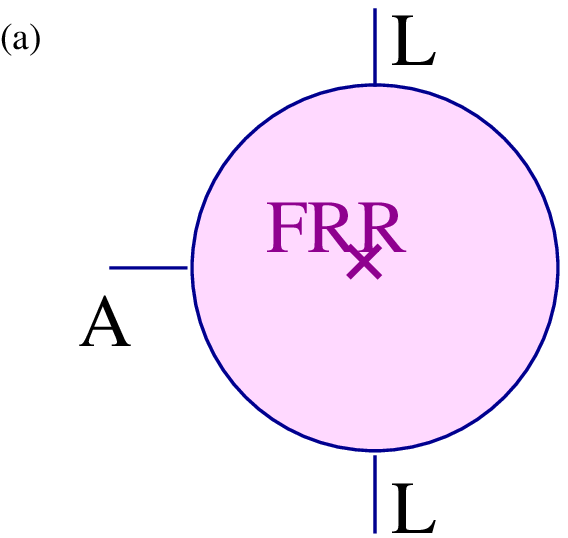}
}
\end{minipage}
\hskip 1cm
\begin{minipage}[b]{0.45\linewidth}
\centering
{
\psfrag{(a)}{\emph{(b)}}
\psfrag{L}{$\Lambda$}
\psfrag{A}{$H_{\mu\nu}$}
\psfrag{FRR}{$\mathcal{F}$}
\includegraphics[width=0.5\textwidth]{LLARR.eps}
}
\end{minipage}
\caption{\small
D3 disk amplitudes involving the R-R background and the gauge field $A_\mu$ \emph{(a)},
the auxiliary field $H_{\mu\nu}$ \emph{(b)}.}
\label{fig:LLHRR}
}
Let us now analyze these mixed open/closed string amplitudes.
When the vertex (\ref{RRvop}) is inserted in the
interior of a disk, the left and right movers of the closed string become identified as
a consequence of the boundary conditions. In the case of a disk
representing the world sheet of a D3 brane, the relevant boundary
conditions for the spin fields are (see, for example, eq. (2.5) of
Ref.~\cite{Billo:2002hm})
\begin{equation}
S^{\dot\alpha}(z)\,S^{(+)}(z) =
{\widetilde S}^{\dot\alpha}(\bar z)\,
{\widetilde{S}}^{(+)}(\bar z)\Big|_{z=\bar z}~,
\label{reflection}
\end{equation}
having conformally mapped the disk to the upper half plane
and hence its boundary to the real axis.
The calculation of a disk amplitude with the
insertion of the closed string vertex (\ref{RRvop}) is then performed by
replacing in the latter the right moving spin fields with the left
moving ones, according to
\begin{equation}
{\widetilde S}^{\dot\alpha}(\bar
z)\,{\widetilde{S}}^{(+)}(\bar z) \,\longrightarrow\,
S^{\dot\alpha}(\bar z)\,S^{(+)}(\bar z) ~~.
\label{reflection1}
\end{equation}
Because of this replacement, any insertion of the R-R vertex (\ref{RRvop}) will
introduce two internal spin fields of type $S^{(+)}$
whose ``charge" has to be compensated by two internal spin fields
of type $S^{(-)}$ in order to have a non-vanishing
amplitude. The only vertex that contains $S^{(-)}$ is
that of the gaugino $\Lambda$ (see eq. (\ref{lambda})), and thus we
easily conclude that any insertion of the graviphoton field
strength $\mathcal{F}_{\dot\alpha \dot\beta}$ must be accompanied by two gauginos
$\Lambda^\alpha$ and $\Lambda^\beta$. However, due to the
different chiralities involved and the symmetry properties
of $\mathcal{F}_{\dot\alpha \dot\beta}$, it is immediate to realize that
some other field is necessary in order to saturate the spinor indices and
produce a non-zero result. Indeed, with only one $V_{\mathcal{F}}$ and
two $V_\Lambda$'s, the correlator among the $\mathrm{SO}(4)$ spin fields
is proportional to $\epsilon_{\alpha\beta}\,\epsilon^{\dot\alpha\dot\beta}$
(see eq. (\ref{4spin_so4}) in appendix \ref{app:sub:correlators})
which vanishes when contracted with the symmetric bi-spinor
$\mathcal{F}_{\dot\alpha \dot\beta}$.
The simplest possibility to avoid this is to insert a gluon vertex
$V_A$, and thus consider the following amplitude
\begin{equation}
\lvev V_{\Lambda}\,V_{\Lambda}\,V_{A}\,V_{\mathcal{F}} \rvev
\equiv C_4\!
\int\frac{\prod_{i} dy_idz d\bar z}{dV_{\rm CKG}}~ \big\langle
V_{\Lambda}(y_1;p_1)\, V_{\Lambda}(y_2;p_2) \,V_{A}(y_3;p_3)
\,V_{\mathcal{F}}(z,\bar z) \big\rangle
\label{LLARR}
\end{equation}
which is represented in Figure~\ref{fig:LLHRR}\emph{a}. Note that the
vertices of the two gauginos and of the graviphoton background already
saturate the superghost charge anomaly, and thus in (\ref{LLARR})
the vertex $V_A$ must be taken in the 0 superghost picture.
In this picture, the properly normalized integrated gluon vertex is (up to ghost terms)
~\cite{Billo:2002hm}
\begin{equation}
\label{vertA0}
{V}_{A}(y;p) = 2\ii\,(2\pi \alpha')^{\frac{1}{2}}\,A_\mu(p)\left(
\partial X^\mu(y) + \ii \,(2\pi \alpha')^{\frac{1}{2}}
\,p\cdot \psi \,\psi^\mu(y)\right)
\ee^{\ii \sqrt{2\pi\alpha'} p \cdot X(y)}
\end{equation}
but, for the reasons explained above, only the
$p\cdot\psi\,\psi^\mu$ part can contribute.
The amplitude (\ref{LLARR}) then becomes
\begin{equation}
\label{LLARR1}
\begin{aligned}
{}&
\!\!\!\!\!\!\!\!
\lvev V_{\Lambda}\,V_{\Lambda}\,V_{A}\,V_{\mathcal{F}} \rvev
=\frac{8}{g_{\rm YM}^2}\,(2\pi\alpha')^{\frac{1}{2}}\,
\mathrm{Tr}\Big(\Lambda^\alpha(p_1)\,\Lambda^\beta(p_2)\,p_3^\nu A^\mu(p_3)\Big)
\,\mathcal{F}_{\dot\alpha \dot\beta}
\\
{}&\times
\int\frac{\prod_{i} dy_i dz d\bar z}{dV_{\rm CKG}}
~\Bigg\{
\big\langle S_\alpha(y_1)\,S_\beta(y_2)\,:\!\psi^\nu\psi^\mu\!:(y_3)\,
S^{\dot\alpha}(z)\, S^{\dot\beta}(\bar z)\big\rangle
\\
{}&\times
\big\langle S^{(-)}(y_1)
\,S^{(-)}(y_2)\,S^{(+)}(z)\,S^{(+)}(\bar z)\big\rangle\,
\big\langle\ee^{-\frac{1}{2}\phi(y_1)}\ee^{-\frac{1}{2}\phi(y_2)}
\ee^{-\frac{1}{2}\phi(z)}\ee^{-\frac{1}{2}\phi(\bar z)}\big\rangle
\\
{}&\times
\big\langle \ee^{\ii \sqrt{2\pi\alpha'}p_1\cdot X(y_1)}
\ee^{\ii \sqrt{2\pi\alpha'}p_2\cdot X(y_2)}
\ee^{\ii \sqrt{2\pi\alpha'}p_3\cdot X(y_3)}\big\rangle \Bigg\}~~.
\end{aligned}
\end{equation}
We now use the correlation functions given in appendix
\ref{app:sub:correlators} and exploit the $\mathrm{SL}(2,\mathbb{R})$
invariance to fix $y_1\to \infty$, $z\to \ii$ and $\bar z \to
-\ii$, so that we are left to perform the following integral%
\footnote{It is also possible to fix the $\mathrm{SL}(2,\mathbb{R})$ symmetry in a more
conventional way by choosing  $y_1=\infty$, $y_2=1$ and $y_3=0$ and
in this way to obtain the integral
$\int_{z\in H^+} dz d\bar z \frac{2\ii y}{|z|^2 \, |1-z|^2}$
over the position of the closed string emission vertex $z=x+\ii y$ in
the upper half plane.
However this integral, as it stands, has a logarithmic divergence for $z\rightarrow
1$; this can be cured by introducing a cutoff
$y>\epsilon$ and letting it go to zero at the end of the computation.
The result we obtain is the same as using the other gauge fixing.
The reason of such a procedure is to avoid that the closed string
emission vertex collides with the border, condition which is
automatically implemented by gauge fixing $z=\ii$.}:
\begin{equation}
\int_{-\infty}^{+\infty}dy_2\,\int_{-\infty}^{y_2}dy_3\,\frac{1}{\left(y_2^2+1\right)
\left(y_3^2+1\right)} = \frac{\pi^2}{2}~~.
\label{integral}
\end{equation}
Collecting all terms, in the end we find
\begin{equation}
\lvev V_{\Lambda}\,V_{\Lambda}\,V_{A}\,V_{\mathcal{F}} \rvev =
\frac{8\pi^2}{g_{\rm YM}^2}\,(2\pi\alpha')^{\frac{1}{2}}\,
\mathrm{Tr}\Big(\Lambda(p_1)\!\cdot\!\Lambda(p_2)\,p_3^\nu A^\mu(p_3)\Big)\,
\mathcal{F}_{\dot\alpha \dot\beta}\left(\bar\sigma_{\nu\mu}\right)^{\dot\alpha \dot\beta}~~.
\label{LLARR2}
\end{equation}
The complete coupling is obtained by multiplying this result by a
symmetry factor of $\frac{1}{2}$ to account for the two alike
gauginos and then by adding to it the amplitude corresponding to
the other inequivalent color order of the three open string vertex
operators; however, these two effects compensate each other and so
the right hand side of (\ref{LLARR2}) is the full answer. {F}rom
this we clearly see that the field theory limit $\alpha'\to 0$
yields a trivial result unless we rescale the graviphoton field
strength $\mathcal{F}_{\dot\alpha \dot\beta}$ to infinity,
in such a way that the following combination
\begin{equation}
4\pi^2\,(2\pi\alpha')^{\frac{1}{2}}\,
\mathcal{F}_{\dot\alpha \dot\beta}\left(\bar\sigma_{\mu\nu}\right)^{\dot\alpha \dot\beta}
\equiv C_{\mu\nu}
\label{cmunu}
\end{equation}
which has dimensions of a (length), remains constant. If we do this, then the amplitude
(\ref{LLARR2}) survives in the field theory limit and produces
the following term in the gauge theory action
\begin{equation}
\frac{\ii}{g_{\rm YM}^2}\int d^4x\,\, \mathrm{Tr}\,\Big(
\Lambda\!\cdot\!\Lambda \big(\partial^\mu A^\nu -\partial^\nu
A^\mu\big)\Big)
C_{\mu\nu}~~.
\label{LLAC}
\end{equation}

Since the gluon vertex operator (\ref{vertA0}) has the same fermionic structure as
 the auxiliary vertex
(\ref{vhmunu}), we should consider also the amplitude
$\lvev V_{\Lambda}\,V_{\Lambda}\,V_{H}\,V_{\mathcal{F}} \rvev$, depicted
in Figure \ref{fig:LLHRR}\emph{b},
whose evaluation follows exactly the same steps we have just
described. In this case we have
\begin{equation}
\lvev V_{\Lambda}\,V_{\Lambda}\,V_{H}\,V_{\mathcal{F}} \rvev =
\frac{2\pi^2}{g_{\rm YM}^2}\,(2\pi\alpha')^{\frac{1}{2}}\,
\mathrm{Tr}\Big(\Lambda(p_1)\!\cdot\!\Lambda(p_2)\,H^{\mu\nu}(p_3)\Big)\,
\mathcal{F}_{\dot\alpha \dot\beta}\left(\bar\sigma_{\nu\mu}\right)^{\dot\alpha
\dot\beta}~~.
\label{LLHRR1}
\end{equation}
Thus, after using (\ref{cmunu}), we conclude that also
the following term must be added to the gauge theory action
\begin{equation}
\frac{1}{2g_{\rm YM}^2}\int d^4x\,\, \mathrm{Tr}\,\Big(
\Lambda\!\cdot\!\Lambda \,H^{\mu\nu}\Big)C_{\mu\nu}~~.
\label{LLHC}
\end{equation}
It is worth pointing out that the disk amplitudes (\ref{LLARR}) and (\ref{LLHRR1})
correspond to 5-point correlation functions from the two-dimensional
world-sheet point of view, since the closed string vertex
$V_{\mathcal{F}}$ effectively counts as two open string vertices
due to the reflection rules (\ref{reflection1}). However, the same amplitudes
correspond to 3-point functions from the point of view of the D3 brane world-volume,
since there are only three vertex operators (those associated to the massless excitations
of the open strings) which carry momentum in four dimensions and
represent dynamical degrees of freedom.

It is not difficult to verify that any other disk amplitude with more
insertions of the R-R vertex operator (\ref{RRvop}), either is zero
because of index structure, or vanishes in the field theory limit
if the combination (\ref{cmunu}) is kept fixed. Thus, even if
we are treating the closed string background in a perturbative
way by means of successive insertions of vertices
$V_{\mathcal{F}}$, in our case this perturbative procedure
terminates after the first step. The terms
(\ref{LLAC}) and (\ref{LLHC}) are then the only two modifications
produced by the graviphoton background in the $\alpha'\to 0$ limit
on the gauge theory action of $N$ D3 branes, which then becomes
\begin{equation}
\label{n22}
\begin{aligned}
{\widetilde S}'&= \frac{1}{g^2_{\rm YM}} \int d^4x ~{\rm
Tr}\Bigg\{\big(\partial_\mu A_\nu-\partial_\nu A_\mu\big)
\partial^\mu A^\nu + 2{\rm i}\, \partial_\mu A_\nu
\big[A^\mu,A^\nu\big]
-\,2\bar\Lambda_{\dot\alpha}\bar
D\!\!\!\!/^{\,\dot\alpha \beta} \Lambda_\beta
\\
& +\ii
 \big(\partial^\mu A^\nu -\partial^\nu A^\mu\big)
\Lambda\!\cdot\!\Lambda\,C_{\mu\nu}
 +H_cH^c +
H_c\,\bar\eta^c_{\mu\nu}\Big(\big[A^\mu,A^\nu\big]+\frac{1}{2}\,\Lambda\!\cdot\!\Lambda
\,C^{\mu\nu}\Big) \Bigg\}~~.
\end{aligned}
\end{equation}
Integrating out the auxiliary field $H$, we finally get
\begin{equation}
\label{n2}
\begin{aligned}
\widetilde S & =
\frac{1}{g^2_{\rm YM}} \int d^4x ~{\rm
Tr}\Bigg\{\frac{1}{2}F_{\mu\nu}^2 -2\bar\Lambda_{\dot\alpha}\bar
D\!\!\!\!/^{\,\dot\alpha \beta} \Lambda_\beta
+\ii\, F^{\mu\nu}\,
\Lambda\!\cdot\!\Lambda\,C_{\mu\nu} -\frac{1}{4}
\,\Big(\Lambda\!\cdot\!\Lambda\,C_{\mu\nu}\Big)^2\Bigg\}
\\
& =
\frac{1}{g^2_{\rm YM}} \int d^4x ~\Tr\Bigg\{
\Big(F_{\mu\nu}^{(-)} + \frac{\ii}{2}
\,\Lambda\!\cdot\!\Lambda\, C_{\mu\nu}\Big)^2 +
\frac{1}{2} F_{\mu\nu}\widetilde F^{\mu\nu} -2\bar\Lambda_{\dot\alpha}\bar
D\!\!\!\!/^{\,\dot\alpha \beta} \Lambda_\beta \Bigg\}
\end{aligned}
\end{equation}
which, in our conventions (see appendix \ref{app:sub:target}), exactly
agrees with the $\mathcal{N}={1}/{2}$ action of \cite{Seiberg:2003yz}.
Therefore, we have shown that the $C$-interactions of
the $\mathcal{N}={1}/{2}$ super Yang-Mills theory, which are usually
derived from a non-anticommutative deformation of the superspace,
can also be obtained directly from string theory, and in particular from the $\alpha'\to 0$ limit
of open string scattering amplitudes in the presence of R-R vertex operators
together with an appropriate rescaling of the graviphoton field strength.

\vspace{0.5cm}
\section{ADHM instanton moduli space in the $\mathcal{N}=1/2$ theory}
\label{sec:D-1}
In this section we describe how the R-R background
that deforms the world-volume dynamics on the D3 branes leading
to the $\mathcal{N}={1}/{2}$ gauge theory, also modifies the moduli space
of its (super)-instantons. Instantons represent an intrinsically non-perturbative
feature of a gauge theory; nevertheless, many aspects of their physics can be reproduced
by perturbative open string computations on systems of D3 branes \emph{and} D-instantons
\cite{Polchinski:fq,Green:1997tv,Dorey:2002ik,Billo:2002hm}.
In this framework, we show that the effects of the graviphoton background on
the D-instantons can be taken into account in a very similar way to what we did in the
previous section for the D3 branes.

\vspace{0.2cm}
\subsection{The undeformed moduli space}
\label{subsec:D-1noRR}
The moduli space of the (super)-instanton solutions of $\mathrm{U}(N)$ (super)-Yang-Mills
theory is described by the ADHM construction \cite{Atiyah:ri}.
This construction can be naturally recast in a stringy language (for a review see, for instance,
\cite{Dorey:2002ik} and references therein); in fact, for instanton number $k$,
one simply adds $k$ D-instantons to the $N$ D3 branes on which the gauge theory lives.
The auxiliary variables appearing in the ADHM construction correspond to the
degrees of freedom of open strings with at least one end-point attached to
a D-instanton.
In \cite{Billo:2002hm} we presented in detail the derivation of the action for the
instanton moduli starting from open string disk amplitudes in flat space,
corresponding to $\mathcal{N}=4$ gauge theory. Here we briefly
review the basic steps of this derivation, adapting it to the $\mathcal{N}=1$ case
with target space $\mathbb{R}^{1,3}\times (\mathbb{R}^6/(\mathbb{Z}_2\times\mathbb{Z}_2))$
which is relevant for our further developments.

In section \ref{sec:D3}, we saw that the tree-level gauge theory action (\ref{n1}) arises
from open string amplitudes computed on disks whose boundaries lie entirely
on the D3 branes, and evaluated in the limit $\alpha' \to 0$
with the coupling $g_{\rm YM}$ and the dimensionful fields $A_\mu$, $\Lambda^\alpha$ and
$\bar\Lambda_{\dot\alpha}$ kept constant.
The moduli action and the ADHM constraints arise instead from open string amplitudes
computed on disks with at least part of their boundaries on the D-instantons.
However, the coupling constant $g_0$ which naturally appears in the ``gauge theory''
on the D-instantons is not independent from $g_{\rm YM}$; the relation between the two
is summarized by writing the normalization $C_0$ of disks attached to
D-instantons~\cite{Billo:2002hm,DiVecchia:1996uq}
\begin{equation}
\label{g0_versus_gym} C_0 =
\frac{1}{2\pi^2{\alpha'}^2}\,\frac{1}{g_{0}^2}
=\frac{8\pi^2}{g_{\rm YM}^2}~~.
\end{equation}
Clearly, if $g_{\rm YM}$ is kept fixed when $\alpha'\to 0$, then $g_0$, which
has dimensions of (length)$^{-2}$, must blow up. This entails the fact that
the moduli have to be rescaled with appropriate powers of $g_0$ to
retain some non-trivial interactions in the field theory limit \cite{Dorey:2002ik,Billo:2002hm}.
In this way, the moduli acquire the dimensions which are appropriate
for their interpretation as parameters of an instanton solution.
For instance, in the NS sector of the
D(--1)/D(--1) strings one would naturally define the ``massless'' vertex operator
\begin{equation}
\label{verta}
V_a(y) = (2\pi\alpha')^{\frac{1}{2}}\,\frac{a_\mu}{\sqrt 2} ~\psi^{
\mu}(y)\,\ee^{-\phi(z)}~~,
\end{equation}
where the moduli $a_\mu$ have dimensions of (length)$^{-1}$, just as
the gluon field $A_\mu$ of the D3/D3 strings.
However, in order to have non-vanishing disk amplitudes in the $\alpha'\to 0$ limit
taken as mentioned above, one must keep fixed the rescaled moduli \cite{Billo:2002hm}
\begin{equation}
\label{rescaled_a} {a'}_\mu = \frac{1}{\sqrt{2} g_0} a_\mu~~,
\end{equation}
which have dimensions of (length) and are related to the position(s) of the
(multi)-centers of the instanton solution.
Note that the above moduli carry also Chan-Paton factors $t^U$ in the
adjoint of $\mathrm{U}(k)$, which are normalized as
\begin{equation}
{\rm tr}\big(t^U\,t^V\big)=\delta^{UV}~~.
\label{normalization2}
\end{equation}

In the R sector of the D(--1)/D(--1) strings on the orbifold, we have four fermionic
moduli ${M'}^\alpha$ and $\lambda^{\dot\alpha}$ which are associated to the vertices
\begin{equation}
\label{vertM'}
\begin{aligned}
V_{M}(y) & = (2\pi\alpha')^{\frac{3}{4}}\,
\frac{g_0}{\sqrt{2}}\,{M'}^{\alpha} \,S_{\alpha}(y)
S^{(-)}(y)\,\ee^{-\frac{1}{2}\phi(y)}~~,
\\
V_{\lambda}(y) & =
(2\pi\alpha')^{\frac{3}{4}}\,{{\lambda'_{\dot\alpha}}}\,S^{\dot\alpha}(y)S^{(+)}(y)
\,\ee^{-\frac{1}{2}\phi(y)}
~~,
\end{aligned}
\end{equation}
where we have already taken into account the rescalings that are suitable
to the $\alpha'\to 0$ limit \cite{Billo:2002hm}. Thus, ${M'}^\alpha$ has
dimensions of (length)$^{\frac{1}{2}}$, while $\lambda'_{\dot\alpha}$ retains
dimensions of (length)$^{-{\frac{3}{2}}}$. Also these moduli have
Chan-Paton factors in the adjoint of $\mathrm{U}(k)$.

Let us now consider the strings that are stretched between a D3
and a D(--1) brane. They are characterized by the fact that the
four longitudinal directions to the D3 branes have mixed boundary
conditions. Thus, in the NS sector of the D3/D(--1) and D(--1)/D3 strings
find the following physical vertices
\begin{equation}
\label{vertexw}
\begin{aligned}
V_w(y) &= (2\pi\alpha')^{\frac{1}{2}}\,
\frac{g_0}{\sqrt{2}}\,{w'}_{\dot\alpha}\, \Delta(y)\,
S^{\dot\alpha}(y) \,\ee^{-\phi(y)}~~,\\
V_{\bar w}(y) &= (2\pi\alpha')^{\frac{1}{2}}\, \frac{g_0}{\sqrt{2}}
\,{\bar w'}_{\dot\alpha}\, \bar\Delta(y)\,
S^{\dot\alpha}(y)\, \ee^{-\phi(y)}~~,
\end{aligned}
\end{equation}
where $\Delta$ and $\bar\Delta$ are twist operators of conformal weight $1/4$
(we refer to appendix \ref{app:sub:correlators} for their
definition and some of their properties).
The bosonic moduli ${w'}_{\dot\alpha}$ and ${\bar w'}_{\dot\alpha}$
carry Chan-Paton factors, respectively,
in the bifundamental representations $\mathbf{N}\times \mathbf{k}$
and $\mathbf{\bar N}\times \mathbf{\bar k}$ of the gauge groups
and therefore one should write more explicitly ${w'}^{iu}_{~~\dot\alpha}$ and
${\bar w'}_{\dot\alpha ui}$, where $u=1,\ldots,N$ and $i=1,\ldots, k$.
As one can see from (\ref{vertexw}), $w'$ and $w'$ have dimensions of a (length) and are
in fact related to the size of the instanton solution.

Finally, in the R sector of the D3/D(--1) and D(--1)/D3 strings, we find the vertices
\begin{equation}
\label{vertexmu}
\begin{aligned}
V_\mu(y) & =
(2\pi\alpha')^{\frac{3}{4}}\, \frac{g_0}{\sqrt{2}}\,{\mu'}\,
\Delta(y)\,S^{(-)}(y)\, \ee^{-{\frac{1}{2}}\phi(y)}~~,
\\
V_{\bar\mu}(y) & =
(2\pi\alpha')^{\frac{3}{4}}\, \frac{g_0}{\sqrt{2}}\,{{\bar \mu'}}\,
\bar\Delta(y)\,S^{(-)}(y)\, \ee^{-{\frac{1}{2}}\phi(y)}~~.
\end{aligned}
\end{equation}
The fermionic moduli $\mu'$ and $\bar \mu'$ have dimensions of
(length)$^{1/2}$, and carry the same Chan-Paton factors as the $w'$'s and $\bar w'$'s.
{F}rom now on, to simplify a bit the notation, we will drop
the primes from all rescaled moduli, except from $a'$ and $M'$
for which they are traditional in the literature.

The vertices (\ref{verta}), (\ref{vertM'}), (\ref{vertexw}) and (\ref{vertexmu})
exhaust the BRST-invariant spectrum of the open strings with at least one
end point on the D-instantons.
However, in order to compute the quartic interactions among the moduli,
it is necessary to introduce \emph{auxiliary moduli} \cite{Billo:2002hm},
which are the strict analogue of the auxiliary fields $H_{\mu\nu}$
we introduced in section \ref{sec:D3} for the D3/D3 gauge theory.
These new auxiliary moduli disentangle the quartic interactions,
so that the moduli action has only cubic terms. The relevant
auxiliary vertex operator that survives the orbifold projection
is
\begin{equation}
\label{vertDmod}
V_D(y) = (2\pi\alpha')\,\frac{{D}_c\,\bar\eta_{\mu\nu}^c}{2}\,
\psi^\nu\psi^\mu(y)~~,
\end{equation}
and describes an excitation of the D(--1)/D(--1) strings.
Note that this vertex is in the 0-superghost picture and that its
polarization has been rescaled according to our general rules
\cite{Billo:2002hm}.

Computing all cubic tree-level interactions among the vertices listed above and taking the
field theory limit (with $g_0\to\infty$) we obtain the following action
for the instanton moduli of the $\mathcal{N}=1$ super Yang-Mills theory
\begin{equation}
\label{smoduli4}
{S}_{\rm mod} = {\rm tr}\Bigg\{\!\!-
\ii D_c\Big({W}^c +\ii
\bar\eta_{\mu\nu}^c \big[{a'}^\mu,{a'}^\nu\big]\Big)\! -
\ii {\lambda}^{\dot\alpha\,} \!\Big({w}^{u}_{~\dot\alpha}\,\bar{\mu}_{u}+
{\mu}^{u}\bar {w}_{\dot\alpha u} +
\big[a'_{\alpha\dot\alpha},{M'}^{\alpha}\big]
\Big)\!
\Bigg\}
\end{equation}
where we introduced the $k\times k$ matrices
\begin{equation}
\label{defWc}
 (W^c)_j^{~i} = w_{~~\dot\alpha}^{iu}\,(\tau^c)^{\dot\alpha}_{~\dot\beta}
\, \bar w^{\dot\beta}_{~uj}
\end{equation}
with $\tau^c$ being the Pauli matrices, and
indicated explicitly the trace over the $\mathrm{U}(k)$ indices $i,j,\ldots$.

The moduli action (\ref{smoduli4}) is much simpler than the corresponding one for the
$\mathcal{N}=4$ theory (see for instance \cite{Dorey:2002ik}) and only accounts for the ADHM
constraints without any further structure. In fact, the moduli
$D_c$ and $\lambda^{\dot\alpha}$ appear as Lagrange multipliers, respectively,
for the bosonic ADHM constraints, which are the following three $k\times k$ matrix equations
\begin{equation}
\label{bosADHM}
{W}^c +\ii
\bar\eta_{\mu\nu}^c\big[{a'}^\mu,{a'}^\nu\big] = \mathbf{0}~~,
\end{equation}
and for their fermionic counterparts
\begin{equation}
\label{fermADHM}
{w}_{~\dot\alpha}^{u}\,\bar{\mu}_{u} +
{\mu}^{u}\bar {w}_{\dot\alpha u} +
\big[a'_{\alpha\dot\alpha},{M'}^{\alpha}\big] = \mathbf{0}~~.
\end{equation}
Once these constraints are satisfied, the moduli action
(\ref{smoduli4}) vanishes.

\vspace{0.2cm}
\subsection{The R-R deformation of the moduli space}
\label{subsec:D-1withRR}
Now we want to take into account the effect on the instanton moduli space of the closed
string R-R background that we introduced in section
\ref{sec:D3}.
To do so we must compute amplitudes on disks which have at least part of their boundary on
the D-instantons, have some insertions of
moduli vertices on the boundary and also some insertions of the
R-R vertex operator (\ref{RRvop}) in the interior of the disk .
In computing these mixed open/closed string amplitudes we must properly take into
account the reflection rules associated to the D(--1) boundary, which relate
the anti-holomorphic to the holomorphic part of the closed vertex operators.
It turns out (see, for example,  eq. (2.4) of Ref. \cite{Billo:2002hm})
that on a D(--1) boundary the spin fields appearing in the R-R
vertex operator (\ref{RRvop}) have exactly the same reflection properties of a D3 boundary
given in (\ref{reflection}). Thus also for the amplitudes we are now considering, we can
replace the right moving parts of the spin fields in the
graviphoton vertex with the left moving ones according to the rule
(\ref{reflection1}).

Let us first consider disks whose boundary lies entirely on the D(--1)
branes; in other words we insert no boundary changing moduli $w$, $\bar w$,
$\mu$ or $\bar \mu$, and hence no twist operators $\Delta$ or $\bar \Delta$.
The situation is then strictly analogous to that of the D3 disks
we considered in section \ref{sec:D3}.
Following the same reasoning given after (\ref{reflection1}), once a R-R
vertex $V_{\mathcal{F}}$ is inserted inside a correlator,
we must insert also two fermionic vertex
operators $V_M$ in order to balance the ``charge'' of the internal spin fields, and
one auxiliary vertex $V_D$ in order to properly saturate the spinor indices and
get a non-vanishing result.
Thus, we must compute the amplitude
\begin{equation}
\lvev V_{M} V_{M} V_D V_{\mathcal{F}} \rvev
\label{MMDF0}
\end{equation}
which corresponds to the diagram depicted in of Figure \ref{fig:mubmuDRR}\emph{a}.
\FIGURE{
\begin{minipage}[b]{0.45\linewidth}
\centering
{
\psfrag{(a)}{\emph{(a)}}
\psfrag{M}{$M'$}
\psfrag{D}{$D_{\mu\nu}$}
\psfrag{FRR}{$\mathcal{F}$}
\includegraphics[width=0.5\textwidth]{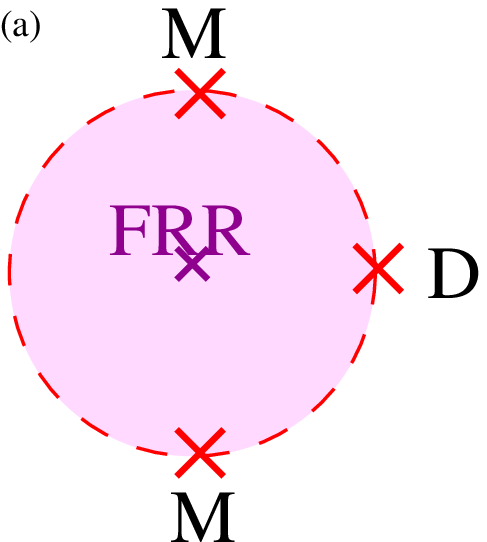}
}
\end{minipage}
\hskip 0.3cm
\begin{minipage}[b]{0.45\linewidth}
\centering
{
\psfrag{(b)}{\emph{(b)}}
\psfrag{mub}{$\bar \mu$}
\psfrag{mu}{$\mu$}
\psfrag{D}{$D_{\mu\nu}$}
\psfrag{FRR}{$\mathcal{F}$}
\includegraphics[width=0.5\textwidth]{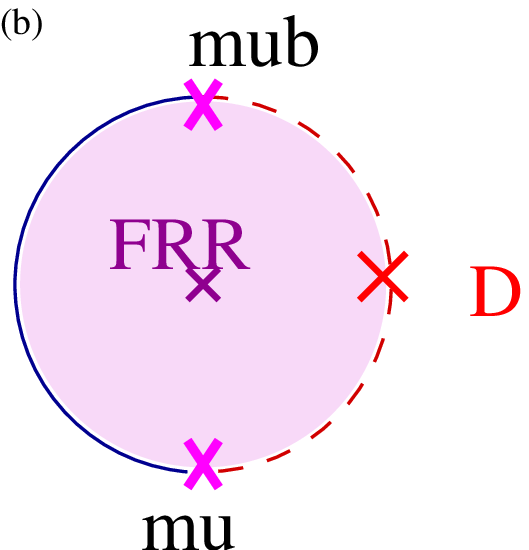}
}
\end{minipage}
\caption{\small Non-zero diagrams with R-R insertions on a D(--1) disk \emph{(a)} and
on a mixed disk \emph{(b)}.}
\label{fig:mubmuDRR}
}
The computation of this amplitude follows exactly
the same steps described for the amplitudes (\ref{LLARR})
and (\ref{LLHRR1}) in section \ref{sec:D3}. Taking into account
the disk normalization $C_0$ given in (\ref{g0_versus_gym}) and the
explicit expressions of the relevant vertices with their proper normalizations, we find
\begin{equation}
\label{risDMMF}
\begin{aligned}
\lvev V_{M} V_{M} V_D V_{\mathcal{F}} \rvev
& = \frac{\pi^2}{2}\,(2\pi\alpha')^{\frac{1}{2}}\,\mathrm{tr}\Big(
M'\!\cdot\! M' D_c\Big)
\bar\eta^c_{\mu\nu}\,\mathcal{F}_{\dot\alpha \dot\beta}
\left(\bar\sigma_{\nu\mu}\right)^{\dot\alpha \dot\beta}
\\
& = -\frac{1}{2}\,\mathrm{tr}\Big(M'\!\cdot\! M' D_c\Big)C^c~,
\end{aligned}
\end{equation}
where we have defined
\begin{equation}
C^c = \frac{1}{4}\bar\eta^c_{\mu\nu}\,C^{\mu\nu}
\label{cc}
\end{equation}
with $C^{\mu\nu}$ being the rescaled graviphoton field-strength
introduced in (\ref{cmunu}).

In the D3/D(--1) system there is also another non-vanishing amplitude involving the
graviphoton background.
Indeed, we can balance the ``charge'' of the internal spin fields of the
R-R vertex $V_{\mathcal{F}}$ also with a pair of boundary changing operators
$V_\mu$ and $V_{\bar \mu}$, so that we should also consider the amplitude
\begin{equation}
\lvev V_{\bar\mu} V_{\mu} V_D V_{\mathcal{F}} \rvev
\label{mumuDF0}
\end{equation}
which corresponds to the \emph{mixed} disk represented in Figure \ref{fig:mubmuDRR}\emph{b}.
At first sight, the evaluation of this mixed amplitude seems rather involved
because the disk has two types of boundary and
hence two types of boundary reflection rules should be implemented on the closed
string vertex operator. However, as we already
mentioned, the spin fields that appear in the
graviphoton vertex (\ref{RRvop}) have the {\it same} boundary
conditions on both kinds of boundaries \cite{Billo:2002hm}, and so also for mixed disks
the reflection properties are those of (\ref{reflection1}).
The amplitude (\ref{mumuDF0}) can then be evaluated following
the same steps described above and using, as specific ingredients,
the correlator of two bosonic twist fields
given in (\ref{deltadelta}) and the $\mathrm{SO}(4)$ correlator among a current
and two spin fields given in (\ref{curr2spinso4}). Taking into account all
normalization factors, in the field theory limit we finally find
\begin{equation}
\label{risDmumuF}
\begin{aligned}
\lvev V_{\bar \mu} V_{\mu} V_D
V_{\mathcal{F}} \rvev
& = \frac{\pi^2}{2} \,(2\pi\alpha')^{\frac{1}{2}}\,
\mathrm{tr}\Big(\bar\mu_u \mu^u D_c\Big) \bar\eta^c_{\mu\nu}\,
\mathcal{F}_{\dot\alpha\dot\beta}
\left(\bar\sigma^{\nu\mu}\right)^{\dot\alpha\dot\beta}
\\
& = -\frac{1}{2}\,\mathrm{tr}\Big(\bar\mu_u \mu^u D_c\Big) C^c~~.
\end{aligned}
\end{equation}
With a systematic analysis one can show that there are no other non-vanishing
diagrams on D(--1) or mixed disks involving the graviphoton
background which survive the $\alpha'\to 0$ limit, and thus
(\ref{risDMMF}) and (\ref{risDmumuF}) are the only terms that modify the
moduli action $S_{\mathrm{mod}}$. Varying such a deformed action
with respect to the auxiliary fields $D_c$, we obtain the modified
ADHM bosonic constraints, which we again write as three $k\times k$ matrix equations
\begin{equation}
\label{bosADHMdef} W^c + \ii \bar\eta^c_{\mu\nu}
\big[{a'}^\mu,{a'}^\nu\big] + \frac{\ii}{2}\Big(M'\!\cdot\! M' - \mu^u \bar\mu_u
\Big) C^c = \mathbf{0}~~.
\end{equation}
Since there are no new types of interactions involving the fermionic
moduli $\lambda^{\dot\alpha}$ and the graviphoton background, the
fermionic ADHM constraints (\ref{fermADHM}) remain unchanged.

We conclude this section by mentioning that a similar analysis can also be
performed to describe the moduli space of anti-instantons ({\it i.e.}
gauge configurations with anti self-dual field strength). In this
case, however, one has to reverse the GSO projections on the
vertex operators of the moduli $w$, $\bar w$, $M'$ and $\lambda$,
which then acquire an opposite $\mathrm{SO}(4)$ chirality
as compared to what we had before, and use an auxiliary vertex
$V_D$ as in (\ref{vertDmod}) but with $\bar\eta^c_{\mu\nu}$
replaced by $\eta^c_{\mu\nu}$. As a consequence of these
changes, any string amplitude involving the anti self-dual
R-R field strength will vanish since the relevant quantity
$C^c$ becomes proportional to $\eta^c_{\mu\nu}\,C^{\mu\nu}$
which is zero. Thus, in the case of anti-instantons the ADHM
constraints are not modified by the anti self-dual graviphoton background; this
result is also in agreement with the structure of the anti-instanton
solutions recently found in \cite{Imaanpur:2003jj,Grassi:2003qk,Britto:2003uv}.

\vspace{0.5cm}
\section{The profile of the deformed instanton solutions}
\label{sec:def_inst}
We now study the instanton solutions of the
$\mathcal{N}=1/2$ gauge theory and analyze how the R-R background affects them.
We adopt the same strategy described in detail in \cite{Billo:2002hm}
where we have shown that the mixed disks of the D3/D(--1) system are the sources for the
classical (super)-instanton solution.
In fact, by computing the emission amplitude for the
gauge vector multiplet from a mixed disk and taking its Fourier transform
after inserting a free propagator, one
obtains the leading term in the large distance expansion of the (super)-instanton
solution in the singular gauge \cite{Billo:2002hm}. For
simplicity, but without loss in generality, here we discuss only the
case of instanton number $k=1$.
\FIGURE{
\begin{minipage}[b]{0.45\linewidth}
\centering
{
\psfrag{(a)}{\emph{(a)}}
\psfrag{I}{\small $I$}
\psfrag{mu}{\small $\mu$}
\psfrag{wb}{\small $\bar w$}
\psfrag{w}{\small $w$}
\psfrag{p}{\small $p$}
\includegraphics[width=0.7\textwidth]{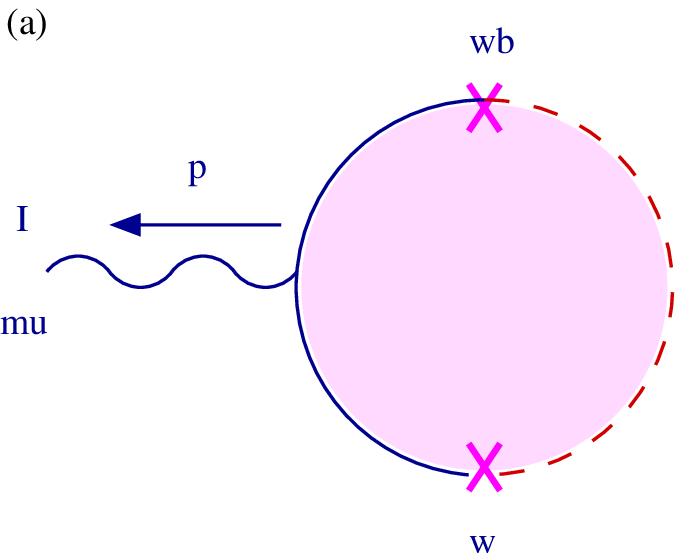}
}
\end{minipage}
\hskip 0.3cm
\begin{minipage}[b]{0.45\linewidth}
\centering
{
\psfrag{(b)}{\emph{(b)}}
\psfrag{I}{$I$}
\psfrag{mu}{$\mu$}
\psfrag{mub}{$\bar \mu$}
\psfrag{mu}{$\mu$}
\psfrag{p}{$p$}
\psfrag{FRR}{$\mathcal{F}$}
\includegraphics[width=0.7\textwidth]{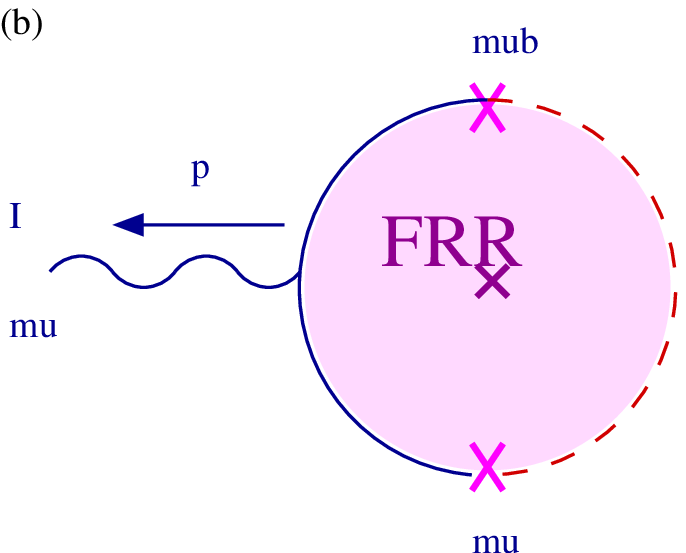}
}
\end{minipage}
\caption{Mixed disks that describe the emission of a gauge vector
field $A_\mu^I$ with momentum $p$ and without a R-R insertion \emph{(a)}
or with a R-R insertion \emph{(b)}.}
\label{fig:asol_RR}
}

Let us begin with the $\mathrm{U}(N)$ gauge field
${A}^I_\mu$. There are two mixed disk diagrams that contribute to the gluon emission
and they are represented in Figures \ref{fig:asol_RR}\emph{a} and \ref{fig:asol_RR}\emph{b}.
The first diagram does not involve the R-R background and
corresponds to the following amplitude
\begin{equation}
\lvev V_{\bar w}\,{\cal V}_{A^I_\mu}(-p)\,V_{w}
\rvev
\label{dia0}
\end{equation}
where in the gluon vertex we have removed the polarization and put
an {\it outgoing} momentum in such a way that the result has the
Lorentz structure and the quantum numbers of an emitted gauge
vector field. Thus, the gluon vertex operator that we must use in
(\ref{dia0}) is (in the 0 superghost picture)
\begin{equation}
\label{vert2}
{\cal V}_{A^I_\mu}(-p) =
2\ii\,(2\pi \alpha')^{\frac{1}{2}}\left(
\partial X^\mu - \ii \,(2\pi \alpha')^{\frac{1}{2}}
\,p\cdot \psi \,\psi^\mu\right)
\ee^{-\ii \sqrt{2\pi\alpha'} p \cdot X}~~.
\end{equation}
As in other amplitudes we have considered before, only the
$p\!\cdot\! \psi \,\psi^\mu$ term contributes in the correlation (\ref{dia0});
performing the calculation we find, as in \cite{Billo:2002hm},
\begin{equation}
\label{corr5}
\lvev V_{\bar w}\,{\cal V}_{A^I_\mu}(-p)\,V_{w}\rvev
= {\ii}\, (T^I)^{v}_{~u}\,p^\nu
\, \bar\eta^c_{\nu\mu}
\big(w^u_{~\dot\alpha}\,(\tau^c)^{\dot\alpha}_{~\dot\beta}
\,\bar w^{\dot \beta}_{~v}\big) \,\ee^{-\ii p\cdot x_0}~.
\end{equation}
where $x_0$ is the location of the D-instanton inside the
world-volume of the D3 branes. Note that all numerical factors and
all powers of $\alpha'$ from the various normalizations completely
cancel out.

We now turn to the second diagram, represented in Figure
\ref{fig:asol_RR}\emph{b}, which instead depends on the R-R
background. It corresponds to the following mixed amplitude
\begin{equation}
\lvev
V_{\bar \mu}\,{\cal V}_{A^I_\mu}(-p)\,V_{\mu}
V_{\mathcal{F}}
\rvev
\label{dia00}
\end{equation}
whose evaluation is identical to that of (\ref{mumuDF0}). Indeed,
we find
\begin{equation}
\label{adacmumu2}
\begin{aligned}
\lvev
V_{\bar \mu}\,{\cal V}_{A^I_\mu}(-p)\,V_{\mu}
V_{\mathcal{F}} \rvev
& = - 2\pi^2\,(2\pi\alpha')^{\frac{1}{2}}\,(T^I)^v_{~u}\, p^\nu
(\bar\sigma_{\nu\mu})^{\dot\alpha\dot\beta}
\,\mathcal{F}_{\dot\alpha\dot\beta}\, \mu^u\bar\mu_v
\,\ee^{-\ii p\cdot x_0}
\\
&=- \frac{1}{2}\,(T^I)^v_{~u}\, p^\nu\, \bar\eta^c_{\nu\mu}\,
\mu^u\bar\mu_v\,C^c\,\ee^{-\ii p\cdot x_0}~~,
\end{aligned}
\end{equation}
where in the last step we have introduced the rescaled graviphoton
field strength according to (\ref{cmunu}) and (\ref{cc}).

There are no other diagrams with only two moduli insertions that
contribute to the emission amplitude of the gauge boson. The latter is
then given by summing (\ref{corr5}) and (\ref{adacmumu2}), namely
\begin{equation}
\label{dia1}
A^I_\mu(p)
= \ii\,  (T^I)^v_{~u}\, p^\nu \, \bar\eta^c_{\nu\mu}
\Big[ (T^c)^u_{~v} + (S^c)^u_{~v} \Big]\,\ee^{-\ii p\cdot x_0}~~,
\end{equation}
where for ease of notation (and for future convenience)
we have introduced the $N\times N$ moduli-dependent matrices
\begin{equation}
\label{STmat}
(T^c)^u_{~v} = w^u_{~\dot\alpha}\,(\tau^c)^{\dot\alpha}_{~\dot\beta}\,
\bar w^{\dot \beta}_{~v}~~~,~~~
(S^c)^u_{~v} = \frac{\ii}{2}\, \mu^u\bar\mu_v\,C^c~~.
\end{equation}

The classical profile of the gauge field in configuration space
is obtained by taking the Fourier transform of the emission
amplitude (\ref{dia1}) after inserting a free propagator, that is
\begin{equation}
\label{gf1}
\begin{aligned}
A_\mu^I(x) &= \int\! \frac{d^4 p}{(2\pi)^2} \,
A_\mu^I(p) \,\frac{1}{p^2}\,\ee^{\ii p\cdot x}
\\
&= 2 \,(T^I)^v_{~u}\, \Big[(T^c)^u_{~v} + (S^c)^u_{~v}\Big]
\,\bar\eta^c_{\mu\nu} \, \frac{(x-x_0)^\nu}{(x-x_0)^4}~~.
\end{aligned}
\end{equation}
As discussed in \cite{Billo:2002hm}, this expression
represents the leading term in the large distance expansion
of the instanton profile. It is important to emphasize that at this stage
the field $A_\mu^I(x)$ in (\ref{gf1}) depends on the \emph{unconstrained}
moduli $w^u_{~\dot\alpha},\bar w^{\dot\beta}_{~v}$, $\mu^u$ and
$\bar\mu_v$ of the ADHM construction, but in
order to get the dependence from the true moduli,
one must enforce the ADHM constraints (\ref{bosADHMdef}) and
(\ref{fermADHM}). In the undeformed theory, if one imposes the
bosonic constraints $W^c=0$, one finds that the matrices $T^c$
generate a $\mathrm{su}(2)$ algebra, while of course the matrices
$S^c$ vanish. Thus, choosing a particular solution of the
constraints amounts simply to choose a particular embedding of an
$\mathrm{SU}(2)$ subgroup inside the gauge group $\mathrm{U}(N)$.
Furthermore, one finds that the gauge field does not have any
component along the $\mathrm{U}(1)$ factor of $\mathrm{U}(N)$.
In this way, from (\ref{gf1}) one retrieves
the large distance behaviour of the standard BPST soliton in the singular
gauge. In the deformed theory, however, the bosonic ADHM constraints imply that
$W^c\not = 0$, and hence these findings are modified.

To see what happens, let us first investigate the algebra of the matrices $T$ and $S$
introduced above. Using their explicit expressions (\ref{STmat}), it is easy to
see that the $S$'s commute among themselves and with the $T$'s, {\it i.e.}
\begin{equation}
\label{commST}
\big[S^a,S^b\big] = 0~~~,~~~
\big[S^a,T^b\big] = 0~~.
\end{equation}
Note that to show the second relation, we must use the fermionic constraint
(\ref{fermADHM}), which for $k=1$ reduces to
$w^u_{~\dot\alpha}\,\bar \mu_u+\varepsilon_{\dot\alpha\dot\beta}\,
\mu^u\,\bar w^{\dot\beta}_{~u} = 0\,$,
actually implying that
\begin{equation}
\label{fermADHMsep}
w_{~\dot\alpha}^{u}\,\bar \mu_u=\mu^u\bar w^{\dot\alpha}_{~u} = 0~~.
\end{equation}
The matrices $T$, with the addition of the matrix
\begin{equation}
\label{T0def}
\big(T^0\big)^u_{~v} =  w^u_{~\dot\alpha}\,\bar w^{\dot \alpha}_{~v}~~,
\end{equation}
are closed under commutation, and satisfy the algebra
\begin{equation}
\label{Talgebra}
\begin{aligned}
\big[T^a,T^b\big] & =  \ii \, \varepsilon^{abc} \left(W^0 T^c - W^c T^0\right)~~,
\\
\big[T^0,T^a\big] & = -\ii \,\varepsilon^{abc} \,W^b T^c~~,
\end{aligned}
\end{equation}
where $W^c \equiv \mathrm{Tr}\big(T^c)$ are exactly the quadratic expressions in the $w$'s
that appear in the ADHM constraint equations (see (\ref{defWc}) for $k=1$),
and $W^0 \equiv \mathrm{Tr}\big(T^0)$.
The algebra (\ref{Talgebra}) can be recast in the form of a
standard $\mathrm{u}(2)$ algebra
\begin{equation}
\label{Thatalgebra}
\big[t^a,t^b\big]  =  \ii \varepsilon^{abc}\, t^c~~~,~~~
\big[t^0,t^a\big]  =  0~~,
\end{equation}
if we define
\begin{equation}
\label{ThatfromT}
\begin{aligned}
t^a & =  \frac{1}{\sqrt{W_0^2 - |\vec W|^2}}
\big( \mathcal{R}^{-\frac{1}{2}}\big)^{ab}
\left(W^0 T^b - W^b T^0\right)~~,
\\
t^0 & = \frac{1}{W_0^2 - |\vec W|^2}\big(W_0 T^0 - \vec W\cdot \vec
T\big)~~,
\end{aligned}
\end{equation}
with $\big(\mathcal{R}\big)^{ab}=W_0^2\,\delta^{ab}-W^aW^b~$.
These generators are normalized in such a way that
$\mathrm{Tr}\big(t^A t^B\big) = \frac{1}{2}
\,\delta^{AB}$ for $A=(0,a)$.
Inverting the above equations we can express the matrices $T$ appearing in the gauge
field profile in terms of the $\mathrm{u}(2)$ generators $t^a$ and $t^0$ as follows
\begin{equation}
\label{TfromThat}
T^a = \mathcal{M}^{ab} t^b + W^b t^0~~,
\end{equation}
where the moduli-dependent matrix $\mathcal{M}$ is
\begin{equation}
\label{explM}
\mathcal{M}^{ab}  = W^0\,\sqrt{W_0^2 - |\vec W|^2}\,
\big( \mathcal{R}^{-\frac{1}{2}}\big)^{ab}~~.
\end{equation}
{F}rom (\ref{commST}) and (\ref{TfromThat}) it follows that the matrices
$S$ commute also with the canonical $\mathrm{u}(2)$ generators,
{\it i.e.} $\comm{S^a}{t^b}=0$ and $\comm{S^a}{t^0}=0\,$.
Using this structure, we can then rewrite the classical solution (\ref{gf1}) as
\begin{equation}
\label{gf2}
A_\mu^I(x) =  2 \Big( \mathcal{M}^{cb}\,\mathrm{Tr}\big(T^I t^b\big)
+ W^c \mathrm{Tr}\big(T^I t^0\big)  +
\mathrm{Tr}\big(T^I S^c\big)\Big)\,\bar\eta^c_{\mu\nu} \, \frac{(x-x_0)^\nu}{(x-x_0)^4}~~.
\end{equation}
{F}rom this result we clearly see that
the $\mathrm{U}(N)$ instanton gauge field contains a part which is aligned, in color space,
along a $\mathrm{U}(2)$ subgroup determined by the $4N$ bosonic moduli
$w_{~\dot\alpha}^u$ and $\bar w^{\dot\alpha}_{~u}$ through the matrices $t^b$ and $t^0$.
Both the non-abelian $\mathrm{SU}(2)$ and the abelian $\mathrm{U}(1)\subset \mathrm{U}(2)$
components are present, in a fashion which is specified by the values of
$W^b$ and $W^0$. Moreover, there is a part of the gauge field along
another abelian factor, commuting with the previous $\mathrm{U}(2)$, that is determined by
the matrices $S^c$ which depend on the fermionic moduli $\mu^u$ and $\bar \mu_u$.
However, to fully specify the instanton profile (including the embedding
of the $\mathrm{U}(2)$ subgroup into $\mathrm{U}(N)$), it is necessary
to take into account the ADHM constraints (\ref{bosADHMdef}) and
(\ref{fermADHM}). For $k=1$, the bosonic ones are just
the following three real equations
\footnote{These three constraints reduce the number of independent bosonic moduli to
$4N -3$. Moreover, a common phase rotation $w\to \ee^{\ii\theta} w, \bar w\to
\ee^{-\ii\theta}\bar w$ leaves invariant the matrices $t^a, t^0$ and their traces.
The true bosonic moduli are therefore $4N-4$, corresponding to the $4N-5$
parameters of the coset $U(N)/(U(N-2)\times U(1))$ plus the size of the instanton.}
\begin{equation}
\label{ADHMbosk=1} W^c =   -\frac{\ii}{2}\Big(M'\!\cdot\! M' - \mu^u \bar\mu_u
\Big) C^c \equiv \hat W^c~~,
\end{equation}
and so all we have to do is simply substitute $W^c=\hat W^c$ in the previous
formulae and obtain the gauge field profile.

To make contact with the $\mathcal{N}=1/2$ instanton solutions recently obtained in
\cite{Grassi:2003qk,Britto:2003uv}, let us choose a specific solution to the
bosonic constraints (\ref{ADHMbosk=1}). Decomposing
the index $u$ as $u=(\dot\beta,i)$, with $\dot\beta=1,2$ and $i=3,\ldots, N$, we set
\begin{equation}
\label{wsol}
\left\{\begin{array}{l}
w^{\dot\beta}_{~\dot\alpha} = \rho \,\delta^{\dot\beta}_{~\dot\alpha} +
\frac{1}{4\rho}\,\hat W^c\,
(\tau^c)^{\dot\beta}_{~\dot\alpha}~~,\cr
w^i_{~\dot\alpha} = 0~~,
\end{array}\right.
\end{equation}
which, in matrix notation, corresponds to choose $w$ as the $N\times 2$ matrix
\begin{equation}
\label{wsol1}
w = \left(\begin{array}{c}
            \rho \,\mathbf{1} + \frac{1}{4\rho}\,\hat W^c \,\tau^c \cr
            \mathbf{0}_{(N-2)\times 2}
          \end{array}\right)~~.
\end{equation}
The moduli $\bar w^{\dot\alpha}_{~u}$ are simply the entries of the hermitian conjugate
matrix $w^\dagger$. It is very easy to verify that with this choice
$W^c \equiv \mathrm{Tr}(w \tau^c w^\dagger) = \hat W^c$~as required;
moreover, the parameter $\rho$ (which, for $\hat W^c=0$, represents the size of the
instanton) appears in
\begin{equation}
\label{wcheck2}
W^0 \equiv \mathrm{Tr}(w w^\dagger) = \frac{1}{2}
\Big(4\rho^2 + \frac{1}{4\rho^2}\,|\vec{\hat W}|^2\Big)~.
\end{equation}
Having fixed $w$ and $\bar w$ as in (\ref{wsol1}), we can make a specific choice of the fermionic
moduli $\mu$ and $\bar \mu$ and solve the constraints (\ref{fermADHMsep}) by setting
\begin{equation}
\label{musol}
\mu^{\dot\alpha} = \bar\mu_{\dot\alpha} = 0~~.
\end{equation}
Furthermore, up to a $\mathrm{U}(N-2)$ rotation, we can choose a single entry of $\mu^i$,
say $\mu^3$, to be different from zero. With this specific choice, we therefore have
\begin{equation}
\label{Wcspec}
\hat W^c = -\frac{\ii}{2}\Big(M'\!\cdot\! M' -\mu^3 \bar\mu_3
\Big) C^c~~,
\end{equation}
and hence expressions of degree three or more in $\hat W^c$ vanish because of the
grassmaniann nature of the parameters $\mu^3$, $\bar\mu_3$ and ${M'}^\alpha$.
All in all, with this specific solution of the ADHM constraints, the
instanton gauge field (\ref{gf2}) can be easily described
by giving its matrix elements $(A_\mu)^u_{~v}$ and decomposing the index
$u$ as $u=(\dot\alpha,i)$, with $i=3,\ldots, N$. The result is
\begin{equation}
\label{Ainu2}
\begin{aligned}
(A_\mu)^{\dot\alpha}_{~\dot\beta} = &
\Bigg\{
\rho^2 (\tau^c)^{\dot\alpha}_{~\dot\beta}
-\frac{\ii}{4}\,\Big(M'\!\cdot\!M'-\mu^3\bar\mu_3\Big)C^c\,\delta^{\dot\alpha}_{~\dot\beta}
\\
& -
\frac{1}{32\rho^2} \Big(|\vec C|^2 (\tau^c)^{\dot\alpha}_{~\dot\beta}
-2 C^c C^b (\tau_b)^{\dot\alpha}_{~\dot\beta}\Big)
M'\!\cdot\! M'\,\mu^3\bar\mu_3\Bigg\}
\,\bar\eta^c_{\mu\nu} \, \frac{(x-x_0)^\nu}{(x-x_0)^4}
\end{aligned}
\end{equation}
for the components in the upper left block. Moreover, there is also
a non-vanishing component outside this block, namely
\begin{equation}
\label{Aoutu2}
(A_\mu)^3_{~3} = \frac{\ii}{2} \,\mu^3 \bar\mu_3\, C_c\,\bar\eta^c_{\mu\nu} \,
\frac{(x-x_0)^\nu}{(x-x_0)^4}~~.
\end{equation}
The above expressions are in agreement with the solution
recently found in \cite{Britto:2003uv}. In
the comparison one has to take into account the different normalizations and conventions,
as well as the fact that their solution is in the regular gauge, while ours is in the
singular gauge.
Furthermore, what we have determined is just the leading term in the long
distance expansion $\rho^2/(x-x_0)^2\ll 1$ of the full instanton
solution.

As discussed in \cite{Billo:2002hm},
mixed disks act as a source also for the gaugino field
$\Lambda_\alpha(x)$. In fact they account for the leading term at long distance of the gaugino
profile in the super-instanton solution
\begin{equation}
\label{gauginosol}
\Lambda^{\alpha,\,I}(x) = -2\ii\,
(T^I)^{v}_{~u} \big(w_{~\dot\beta}^{u}\,{\bar \mu}_{v}+
\mu^{u}\,{\bar w}_{\dot\beta v}\big)
(\bar\sigma_\nu)^{\dot\beta\alpha} \frac{(x - x_0)^\nu}{(x -
x_0)^4}
+ \frac{\ii}{2}\,
{M'}^{\beta}\,(\sigma^{\mu\nu})_{\beta}^{~\alpha}\,
F^I_{\mu\nu}(x)
\end{equation}
where $F^I_{\mu\nu}$ is the gauge field strength. No diagram involving
R-R insertions that could correct this result survives in the field theory
limit, and thus (\ref{gauginosol}) is the gaugino profile at large distance
also in the $\mathcal{N}=1/2$ theory. Finally, we recall that with
the replacement
\begin{equation}
{M'}^\alpha \longrightarrow {M'}^\alpha - \bar\zeta_{\dot\alpha}
(\bar\sigma^{\mu})^{\dot\alpha \beta}\,a'_\mu
\label{superconf}
\end{equation}
in all previous formulas one can account for the superconformal zero-modes of the instanton
that are parameterized by $\bar \zeta$.

We conclude by noting that the sub-leading terms in the large distance expansion
of the super-instanton solution can be obtained by a perturbative
analysis \cite{Billo:2002hm} in which more sources ({\it i.e.} more mixed disks)
emit each a gauge boson or a gaugino, which then interact with the (deformed) vertices
of the $\mathcal{N}=1/2$ Yang-Mills theory to produce a single gauge boson or gaugino.
However, this is exactly the same procedure which has been followed in
\cite{Grassi:2003qk,Britto:2003uv} to determine in a purely
field-theoretical framework the (deformed) super-instanton solution,
and hence to repeat it here would not add much to our discussion.
On the other hand, in the evaluation of instanton-induced or
instanton-modified correlators one typically takes into account just
the leading contribution in the large-distance expansion of the instanton
solution in the singular gauge, which is what the mixed disks provide.

It would be interesting to generalize this analysis
to models with extended supersymmetry and to other
kinds of closed string backgrounds. It would be nice also to
repeat the calculation of the open string scattering amplitudes
presented in this paper using the Berkovits formalism \cite{Berkovits:2002zk} which,
in contrast to the RNS formalism, allows
to treat the R-R background in an exact manner.

\vskip 1.5cm
\noindent {\large {\bf Acknowledgments}}
\vskip 0.2cm
\noindent We thank Pietro Antonio Grassi, Stefano Sciuto, Giuseppe Vallone,
the organizers and the participants of the workgroup on $\mathcal{N}=1/2$ gauge
theories held at the
``RTN 2004 Winter School on Strings, Supergravity and Gauge Theory'', Barcelona, 12-16 Jan. 2004,
for fruitful discussions.
We are especially grateful to Francesco Fucito whose collaboration on related subjects
helped us to tackle this problem.

\vspace{0.5cm}
\appendix

\section{Notations and conventions}
\label{app:conventions}

\vspace{0.2cm}
\subsection{Target-space conventions}
\label{app:sub:target}

\paragraph{Indices:}
We denote by $\mu=1,2,3,4$ the directions in the 4-dimensional Euclidean world-volume
of the D3 branes. By $\alpha$ and $\dot\alpha$ we denote, respectively, chiral and
anti-chiral spinor indices in the same space. We use $u,v,\ldots=1,\ldots,N$ to
enumerate the D3-branes and $i,j=1,\ldots,k$ to enumerate the D-instantons.
The indices $u,v,\ldots$ transform in the fundamental (or anti-fundamental,
depending whether they are in upper or lower position) of the $\mathrm{U}(N)$ gauge group,
while the indices $i,j,\ldots$ transform
in the (anti)-fundamental of $\mathrm{U}(k)$.
We reserve capital indices $I,J,\ldots$ for the adjoint of $\mathrm{U}(N)$.

\paragraph{Gauge fields:}
We define the non-abelian field strength in terms of a hermitian connection
$A_\mu = A_\mu^I\, T^I$ as
\begin{equation}
\label{deffs}
F_{\mu\nu} = \partial_\mu A_\nu -\partial_\nu A_\mu +\ii \comm{A_\mu}{A_\nu}~.
\end{equation}

\paragraph{$\mathbf{d=4}$ Clifford algebra:}
Let us define the matrices $(\sigma^\mu)_{\alpha\dot\beta}$ and
$(\bar\sigma^{\mu})^{\dot\alpha\beta}$
with
\begin{equation}
\label{sigmas}
\sigma^\mu =
(\ii\vec\tau,\mathbf{1})~~~,~~~
\bar\sigma^\mu =
\sigma_\mu^\dagger = (-\ii\vec\tau,\mathbf{1})~~,
\end{equation}
where $\tau^c$ are the ordinary Pauli matrices. They satisfy
the Clifford algebra
\begin{equation}
\label{cliff4}
\sigma_\mu\bar\sigma_\nu + \sigma_\nu\bar\sigma_\mu =
2\delta_{\mu\nu}\,\mathbf{1}~~,
\end{equation}
and correspond to a Weyl representation of the $\gamma$-matrices
acting on chiral or anti-chiral spinors $\psi_\alpha$ or $\psi^{\dot\alpha}$.
Out of these matrices, the $\mathrm{SO}(4)$ generators are defined by
\begin{equation}
\label{sigmamunu}
\sigma_{\mu\nu}
=\frac{1}{2}(\sigma_\mu\bar\sigma_\nu -
\sigma_\nu\bar\sigma_\mu)~~~,~~~
\bar\sigma_{\mu\nu}
=\frac{1}{2}(\bar\sigma_\mu\sigma_\nu -
\bar\sigma_\nu\sigma_\mu)~~.
\end{equation}
The matrices $\sigma_{\mu\nu}$ are self-dual and thus generate the
$\mathrm{SU}(2)_\mathrm{L}$ factor of $\mathrm{SO}(4)$;
the anti self-dual matrices $\bar\sigma_{\mu\nu}$ generate
instead the $\mathrm{SU}(2)_\mathrm{R}$ factor.
The charge conjugation matrix $C$ is block-diagonal in the Weyl basis,
and is given by $C^{\alpha\beta} = -\varepsilon^{\alpha\beta}$ and
$C^{\dot\alpha\dot\beta} = -\varepsilon^{\dot\alpha\dot\beta}$
with $\varepsilon^{12}=\varepsilon_{12}
=-\varepsilon^{\dot 1\dot 2}=-\varepsilon_{\dot 1\dot 2}=+1$.
Moreover we raise and lower spinor indices as follows
\begin{equation}
\psi^\alpha=\varepsilon^{\alpha\beta}\,\psi_\beta
~~~,~~~
\psi_{\dot\alpha}= \varepsilon_{\dot\alpha\dot\beta}\,\psi^{\dot\beta}~~.
\end{equation}
The generators $(\sigma^{\mu\nu})_{\alpha\beta}$ and
$(\bar\sigma^{\mu\nu})^{\dot\alpha\dot\beta}$, in which
the indices have been lowered or raised according to the above rule,
are \emph{symmetric} in the spinor indices.

The explicit mapping of a self-dual $\mathrm{SO}(4)$
tensor into the adjoint representation of the $\mathrm{SU}(2)_\mathrm{L}$ factor
is realized by the 't Hooft symbols $\eta^c_{\mu\nu}$; the analogous
mapping of an anti self-dual tensor into the adjoint
of the $\mathrm{SU}(2)_\mathrm{R}$ subgroup is realized by
$\bar\eta^c_{\mu\nu}$. Specifically we have
\begin{equation}
\label{sigmaeta}
(\sigma_{\mu\nu})_{\alpha}^{~\beta} =
\ii \,\eta^c_{\mu\nu}\, (\tau^c)_{\alpha}^{~\beta}
~~~,~~~
(\bar\sigma_{\mu\nu})^{\dot\alpha}_{~\dot\beta} = \ii \,\bar\eta^c_{\mu\nu}\,
(\tau^c)^{\dot\alpha}_{~\dot\beta}~~.
\end{equation}
Interpreted as $4\times 4$ matrices, the 't Hooft symbols satisfy the algebra
\begin{equation}
\label{quat_algebra}
\eta^c\eta^d = - \delta^{cd}\mathbf{1} - \varepsilon^{cde}\eta^e
\end{equation}
with an analogous formula for the $\bar\eta$'s.
We also have
\begin{eqnarray}
\eta^c_{\mu\nu}\,\eta^{d\,\mu\nu} &=& 4\,\delta^{cd}~~,\\
\eta^{c}_{\mu\nu}\,\eta^{c}_{\rho\sigma}&=&
\delta_{\mu\rho}\,\delta_{\nu\sigma}
-\delta_{\mu\sigma}\,\delta_{\nu\rho}\,+\,
\varepsilon_{\mu\nu\rho\sigma}~~.
\label{etaeta}
\end{eqnarray}
Analogous formulas hold for the $\bar\eta$'s
with a minus sign in the $\varepsilon$ terms of (\ref{etaeta}).
{F}rom (\ref{etaeta}) and (\ref{sigmaeta}) it also follows
\begin{equation}
\label{sigmasigma}
\begin{aligned}
\tr \big(\sigma^{\mu\nu} \sigma^{\rho\sigma}\big)
& =
2 \big(\delta_{\mu\rho}\delta_{\nu\sigma} - \delta_{\mu\sigma}\delta_{\nu\rho} +
\varepsilon_{\mu\nu\rho\sigma}\big)~~,
\\
\tr \big(\bar\sigma^{\mu\nu} \bar\sigma^{\rho\sigma}\big)
& =
2 \big(\delta_{\mu\rho}\delta_{\nu\sigma} - \delta_{\mu\sigma}\delta_{\nu\rho} -
\varepsilon_{\mu\nu\rho\sigma}\big)~~,
\end{aligned}
\end{equation}
where the trace is over the undotted or dotted spinor indices.
Another useful formula is
\begin{equation}
\label{tauctauc}
(\tau_c)^{\dot\alpha}_{~\dot\beta} (\tau^c)^{\dot\gamma}_{~\dot\delta} =
\delta^{\dot\alpha}_{~\dot\delta} \delta^{\dot\gamma}_{~\dot\beta} -
\varepsilon^{\dot\alpha\dot\gamma} \varepsilon_{\dot\beta\dot\delta}~~,
\end{equation}
from which, after using (\ref{sigmaeta}), it follows
\begin{equation}
\label{tauctaucup}
(\bar\sigma^{\mu\nu})^{\dot\alpha\dot\beta}
(\bar\sigma_{\mu\nu})^{\dot\gamma\dot\delta}
= -4
(\tau_c)^{\dot\alpha\dot\beta} (\tau^c)^{\dot\gamma\dot\delta} =
4 (\varepsilon^{\dot\alpha\dot\gamma} \varepsilon^{\dot\beta\dot\delta}
+ \varepsilon^{\dot\alpha\dot\delta} \varepsilon^{\dot\beta\dot\gamma} )~~.
\end{equation}

\paragraph{(Anti) self-dual tensors:}
Any antisymmetric tensor $\mathcal{F}_{\mu\nu}$ decomposes into a self-dual
and an anti self-dual component according to $\mathcal{F}_{\mu\nu} =
\mathcal{F}^{(+)}_{\mu\nu} + \mathcal{F}^{(-)}_{\mu\nu}$ where
\begin{equation}
\label{defasd}
\mathcal{F}_{\mu\nu}^{(\pm)} = \pm\frac{1}{2} \varepsilon_{\mu\nu\rho\sigma}
\mathcal{F}^{(\pm)}_{\rho\sigma} ~~.
\end{equation}
We can also write $\mathcal{F}_{\mu\nu}^{(\pm)} = (\mathcal{F}_{\mu\nu}
\pm \widetilde{\mathcal{F}}_{\mu\nu})/2$,
with $\widetilde{\mathcal{F}}_{\mu\nu}\equiv  \varepsilon_{\mu\nu\rho\sigma}
{\mathcal{F}}^{\rho\sigma}/2$~.

Given an anti self-dual tensor $\mathcal{F}_{\mu\nu}^{(-)}$,
we can map it to a 3-vector transforming in the adjoint representation
of $\mathrm{SU}(2)_\mathrm{R}$ using the anti self-dual t'Hooft symbols
$\bar \eta^c_{\mu\nu}$ according to
\begin{equation}
\label{defFc}
\mathcal{F}_{\mu\nu}^{(-)} = \mathcal{F}_c\bar\eta^c_{\mu\nu}~~~,~~~
\mathcal{F}^c = \frac{1}{4} \mathcal{F}^{(-)\mu\nu}
\bar\eta^c_{\mu\nu}~~.
\end{equation}
We can organize the three degrees of freedom
of the anti self-dual tensor into a \emph{symmetric} dotted bi-spinor
by setting
\begin{equation}
\label{defFab}
\mathcal{F}_{\mu\nu}^{(-)} = \frac{1}{2}\, \mathcal{F}_{\dot\alpha\dot\beta}
(\bar\sigma^{\mu\nu})^{\dot\alpha\dot\beta}~~~,~~~
\mathcal{F}_{\dot\alpha\dot\beta} = \frac{1}{4}\, \mathcal{F}_{\mu\nu}^{(-)}
(\bar\sigma^{\mu\nu})_{\dot\alpha\dot\beta}~~.
\end{equation}
Using (\ref{sigmaeta}), we can also write
\begin{equation}
\label{FcandFab}
\mathcal{F}^c =  \frac{\ii}{2} \,\mathcal{F}_{\dot\alpha\dot\beta}
(\tau^c\varepsilon)^{\dot\alpha\dot\beta}~~~,~~~
\mathcal{F}_{\dot\alpha\dot\beta}  =  \ii \,\mathcal{F}_c
(\varepsilon\tau^c)_{\dot\alpha\dot\beta}~~.
\end{equation}
Given any two anti self-dual tensors $\mathcal{F}_{\mu\nu}^{(-)}$ and
$\mathcal{G}_{\mu\nu}^{(-)}$, we can contract them as follows
\begin{equation}
\label{contr_asd}
\vec{\mathcal{F}}\cdot\vec{\mathcal{G}} \equiv \mathcal{F}^c\mathcal{G}_c
=\frac{1}{4}\,\mathcal{F}^{(-)\mu\nu}\mathcal{G}_{\mu\nu}^{(-)} = \frac{1}{2}\,
\mathcal{F}^{\dot\alpha\dot\beta}
\mathcal{G}_{\dot\alpha\dot\beta}~~,
\end{equation}
where in the last step we have used (\ref{sigmasigma}).

\paragraph{The internal orbifold space:}
In order to engineer a $\mathcal{N}=1$ gauge theory with D3-branes,
we take the six-dimensional transverse space to be an orbifold,
obtained by modding out the space $\mathbb{R}^6$
corresponding to the directions $x^5,\ldots,x^{10}$ (and to $\psi^5,\ldots\psi^{10}$)
by the action of a $\mathbb{Z}_2\times \mathbb{Z}_2$ group.
The two generators $g_1$ and $g_2$ of
this group act as follows: $g_1$ is a $\pi$ rotation in the 7-8 plane and a $-\pi$ rotation in
the 9-10 plane; $g_2$ is a $\pi$ rotation in the 5-6 plane and a $-\pi$ rotation
in the 9-10 plane.

Given the Clifford algebra of the matrices $\gamma^5,\ldots \gamma^{10}$ (which in our stringy
perspective are related to the 0-modes of $\psi^5,\ldots
\psi^{10}$), one can easily see that
the combinations $e^\pm_{5-6} = (\gamma^5 \pm \ii \gamma^6)/2$~,
$e^\pm_{7-8} = (\gamma^7 \pm \ii \gamma^8)/2$ and
$e^\pm_{9-10} = (\gamma^9 \pm \ii \gamma^{10})/2$
are fermionic creation and annihilation operators.
Thus, the 8-dimensional spinor space is spanned by the states $\ket A  =
\ket{\pm \frac{1}{2}}_{5,6}\otimes\ket{\pm \frac{1}{2}}_{7,8}\otimes\ket{\pm \frac{1}{2}}_{9,10}$,
where $\ket{\pm \frac{1}{2}}_{5,6}$ have eigenvalues $\pm {\ii}/{2}$ with respect to the
the Lorentz generator $J_{56} = \comm{\gamma^5}{\gamma^6}/4 = \ii\sigma^3/2$, and similarly
for the 7-8 and 9-10 directions. Then, on the spinor space the two generators of the
$\mathbb{Z}_2\times \mathbb{Z}_2$ group are
\begin{equation}
\label{spinorbifoldg1}
\begin{aligned}
g_1 &\to \mathbf{1}\otimes\ee^{\pi J_{78}}\otimes\ee^{-\pi J_{9,10}} =
\mathbf{1}\otimes\ee^{\ii{\frac{\pi}{2}}\sigma^3}\otimes\ee^{-\ii{\frac{\pi}{2}}\sigma^3} =
\mathbf{1}\otimes (\ii \sigma^3) \otimes (-\ii \sigma^3)~~,
\\
g_2 &\to \ee^{\pi J_{56}}\otimes \mathbf{1}\otimes\ee^{-\pi J_{9,10}} =
\ee^{\ii{\frac{\pi}{2}}\sigma^3}\otimes\mathbf{1}\otimes\ee^{-\ii{\frac{\pi}{2}}\sigma^3}
= (\ii \sigma^3) \otimes \mathbf{1}  \otimes (-\ii \sigma^3)~~.
\end{aligned}
\end{equation}
It is easy to see that the only spinor states which are invariant under $g_1$
and $g_2$ are $\ket{+\frac{1}{2}}_{5,6}\otimes
\ket{+\frac{1}{2}}_{7,8}\otimes \ket{+\frac{1}{2}}_{9,10}$
and $\ket{-\frac{1}{2}}_{5,6}\otimes \ket{-\frac{1}{2}}_{7,8}
\otimes \ket{-\frac{1}{2}}_{9,10}$.
In other words, the only surviving spinor weights are
\begin{equation}
\label{orbi_spin_weights}
\vec\lambda^{(+)} = (+\frac{1}{2},+\frac{1}{2},+\frac{1}{2})~~~,
~~~
\vec\lambda^{(-)} = (-\frac{1}{2},-\frac{1}{2},-\frac{1}{2})~~.
\end{equation}
The first one is chiral, whilst the second is anti-chiral.

\vspace{0.2cm}
\subsection{World-sheet conventions}
\label{app:sub:correlators}

\paragraph{Spin fields and bosonization:} As usual, to discuss $\mathrm{SO}(2N)$ spin
fields we utilize the Frenkel-Ka\c c \cite{Frenkel-Kac} construction (see, for example,
\cite{Kostelecky:1986xg}).
Out of $2N$ world-sheet fermions $\psi^m$, a $\mathrm{SO}(2N)$ current
is defined as $J^{mn} = :\!\psi^m\psi^n\!:$~. Grouping the directions in pairs,
one introduces $N$ world-sheet bosons $\varphi_i$, ($i=1,\ldots N$) by
\begin{equation}
\label{bosonizdef}
\frac{\psi^{2i-1} \pm\ii \,\psi^{2i}}{\sqrt{2}} = c_i~\ee^{\pm \ii\,\varphi_i}~~,
\end{equation}
where $c_i$ are cocycle factors needed to maintain the fermionic statistic.
In a Cartan basis with Cartan generators $H_i = J^{2i-1,2i} $ $
= :\!\psi^{2i-1} \psi^{2i}\!:$~, from (\ref{bosonizdef}) we get
\begin{equation}
\label{cartanbos}
H_i = \ii\partial \varphi_i~~.
\end{equation}
Generators associated to a root $\vec\alpha$ are represented by $E_{\vec\alpha}
= \ee^{\ii\vec\alpha\cdot\vec\varphi}$; more generally, operators transforming
under $\mathrm{SO}(2N)$ as specified by a weight vector $\vec\lambda$ are realized as
\begin{equation}
\label{weightbos}
O_{\vec\lambda}=  c_{\vec\lambda}~\ee^{\ii\vec\lambda\cdot\vec\varphi}~~,
\end{equation}
where again $c_{\vec\lambda}$ is a cocycle factor.

Spin fields transform in a spinor representation: $S^A$ is associated to a spinor weight
$\vec\lambda^A$, with $\lambda^A_i = \pm \frac{1}{2}$ (if the product of all the signs is
plus or minus, the spinor is, respectively, chiral or anti-chiral).

Correlators among operators of definite $\mathrm{SO}(2N)$ weights are easily found
in the bosonized formulation, since for each boson $\varphi_i$ we have
\begin{equation}
\label{corbos}
\Big\langle\prod_{k} \ee^{\ii \beta^k \varphi_i}(y_k)\Big\rangle
\simeq
\delta\big(\sum_k \beta^k\big)\,\prod_{k<m} (y_k-y_m)^{-\beta_k
\beta_m}~~,
\end{equation}
and other well-known formulae when also $\partial \varphi_i$ operators are inserted.

In deriving the correlators listed below by means of the
bosonization formulae, we will not explicitly take into account
the cocycle factors, but rather summarize
their presence into ``effective'' rules for the choice of signs and phases.

\paragraph{Spacetime $\mathbf{{\rm SO}(4)}$ correlators:}
Our bosonization conventions are that the chiral spin fields $S^{\alpha}$ correspond
to the weights $(+\frac{1}{2},+\frac{1}{2})$ for $\alpha=1$ and
$(-\frac{1}{2},-\frac{1}{2})$ for $\alpha=2$. For the anti-chiral
spin fields $S^{\dot\alpha}$, instead, $\dot\alpha=1$ corresponds to $(+\frac{1}{2},-\frac{1}{2})$
and $\dot\alpha=2$ to $(-\frac{1}{2},+\frac{1}{2})$. With these positions,
and the general formulae discussed above, one derives the following correlators
that have been used in the main text.

The non-vanishing 4-point correlator involving spin fields of different chiralities
is
\begin{equation}
\label{4spin_so4}
\big\langle S_\gamma(y_1) S_\delta(y_2)
S^{\dot\alpha}(z) S^{\dot\beta}(\bar z) \big\rangle  =\varepsilon_{\gamma\delta}
\,\varepsilon^{\dot\alpha\dot\beta} \,(y_1-y_2)^{-\frac{1}{2}}\,
(z-\bar z)^{-\frac{1}{2}}~~,
\end{equation}
while the correlator with a current and two spin fields
is given by
\begin{equation}
\label{curr2spinso4}
\big\langle
:\!\psi_\mu\psi_\nu\!:\!(y_3) \,S^{\dot\alpha}(z) S^{\dot\beta}(\bar z)
\big\rangle =
\frac{1}{2} \,
(\bar\sigma_{\mu\nu})^{\dot\alpha\dot\beta}
(z-\bar z)^{\frac{1}{2}}\,(y_3-z)^{-1}\,(y_3-\bar z)^{-1}~~.
\end{equation}
A similar formula holds for chiral spin fields.
A 5-point correlators between one current and four spin fields plays a crucial
role in the present paper and is given by
\begin{equation}
\label{curr4spin_so4}
\begin{aligned}
{}&  \big\langle S_\gamma(y_1) S_\delta(y_2)
:\!\psi^\mu\psi^\nu\!:\!(y_3)\,
S^{\dot\alpha}(z) S^{\dot\beta}(\bar z) \big\rangle
=
\frac{1}{2}\,(y_1-y_2)^{-\frac{1}{2}}(z-\bar z)^{-\frac{1}{2}}
\\
{}& ~~\times
\Bigg(
(\sigma^{\mu\nu})_{\gamma\delta}\,
\varepsilon^{\dot\alpha\dot\beta}\,
\frac{(y_1-y_2)}{(y_1 - y_3)(y_2 - y_3)} \
+ \varepsilon_{\gamma\delta}
\,(\bar\sigma^{\mu\nu})^{\dot\alpha\dot\beta}\,
\frac{(z-\bar z)}{(y_3 - z)(y_3 - \bar z)}
\Bigg)~~.
\end{aligned}
\end{equation}

\paragraph{Correlators on $\mathbb{R}^6/(\mathbb{Z}_2\times \mathbb{Z}_2)$:}
According to (\ref{orbi_spin_weights}) the only surviving spin fields on the orbifold
$\mathbb{R}^6/(\mathbb{Z}_2\times \mathbb{Z}_2)$ are
\begin{equation}
\label{orbi_spin_fields}
S^{(+)} = \ee^{\frac{\ii}{2}(\varphi_1 + \varphi_2 +
\varphi_3)}~~~,~~~
S^{(-)} = \ee^{-\frac{\ii}{2}(\varphi_1 + \varphi_2 + \varphi_3)}~~,
\end{equation}
up to cocycle factors.
There is a single non-vanishing 4-spin correlator, which is crucial in our computations
and is given by
\begin{equation}
\label{4spin_so6}
\begin{aligned}
{}&\!\!\!\!
\big\langle S^{(-)}(y_1) S^{(-)}(y_2) S^{(+)}(z)
S^{(+)}(\bar z) \big\rangle
\\
{}&~~ = (y_1-y_2)^{\frac{3}{4}}\,
(y_1-z)^{-\frac{3}{4}}\, (y_1-\bar z)^{-\frac{3}{4}} \, (y_2-z)^{-\frac{3}{4}}\,
(y_2-\bar z)^{-\frac{3}{4}} \,(z-\bar z)^{\frac{3}{4}}~~.
\end{aligned}
\end{equation}

\paragraph{Bosonic twist fields:}
For the D3/D(--1) and the D(--1)/D3 strings, the fields $X^\mu$
along the world-volume of the D3 branes
describe Neumann-Dirichlet directions. Their twisted boundary conditions
can be seen as due to twist and anti-twist fields $\Delta$ and
$\bar\Delta$ that change the boundary conditions from Neumann to
Dirichlet and vice-versa by introducing a cut in the world-sheet
(see for example Ref.~\cite{Dixon:jw}).
The twist fields $\Delta$ and $\bar\Delta$ are bosonic
operators with conformal dimension $1/4$ and their OPE's are
\begin{equation}
\Delta(y_1)\,\bar\Delta(y_2) \sim (y_1-y_2)^{-\frac{1}{2}}~~~,~~~
\bar\Delta(y_1)\,\Delta(y_2) \sim -\,(y_1-y_2)^{-\frac{1}{2}} ~~,
\label{deltadelta}
\end{equation}
where the minus sign in the second correlator is  an ``effective''
rule to correctly account for the space-time statistics in correlation
functions. More generally, one can show that
\begin{equation}
\label{corr2}
\left\langle \,\bar \Delta(y_1) \,\ee^{-\ii \sqrt{2\pi\alpha'} p\cdot
X(y_2)} \,\Delta(y_3)\,\right\rangle = -\,\ee^{-\ii p\cdot x_0}\,
(y_1-y_3)^{-\frac{1}{2}}~~,
\end{equation}
where $x_0$ denotes the location of the D-instantons inside the world-volume
of the D3 branes. This correlator is crucial in computing the profile of the fields
emitted by mixed disks, as shown in (\ref{corr5}) and (\ref{adacmumu2}).

\paragraph{Superghosts:}
As usual, we adopt the bosonized treatment of \cite{Friedan:1985ge} of the superghost system.
We use systematically the following correlator between vertices of the type
$\ee^{-\frac{1}{2}\phi}$, where $\phi$ is the chiral boson (with background charge
2) introduced
in this formalism, namely
\begin{equation}
\label{4sghostcorr}
\begin{aligned}
{}&\!\!\!\!
\big\langle\ee^{-\frac{1}{2}\phi(y_1)}\,\ee^{-\frac{1}{2}\phi(y_2)}\,
\ee^{-\frac{1}{2}\phi(z)}\,
\ee^{-\frac{1}{2}\phi(\bar z)}\big\rangle
\\
{}&~~= \Big[(y_1-y_2)\,(y_1-z)\,(y_1-\bar z)\,(y_2 - z)\,
(y_2 - \bar z)\,(z - \bar z)\Big]^{-\frac{1}{4}}~~.
\end{aligned}
\end{equation}

\paragraph{Conjugation conventions:}
In the NS sector, the conjugation properties of the polarizations are unambiguously
fixed by the expression of the associated vertices themselves.
As an example, consider the vertices for the $w$ and $\bar w$ moduli, given in
(\ref{vertexw}). The conjugate of $\Delta\,S^{\dot\alpha}\,\ee^{-\phi}$
is determined by the two-point functions of the involved conformal fields, and
is $\bar\Delta \,S_{\dot\alpha}\,\ee^{-\phi}$. {F}rom this fact, we deduce the following
conjugation rule
\begin{equation}
\label{conjw}
(w^{iu}_{~~\dot\alpha})^* = \bar w^{\dot\alpha}_{~ui}~,
\end{equation}
or simply $(w_{\dot\alpha})^\dagger = \bar w^{\dot\alpha}$ in a $k\times N$ matrix
notation for $w_{\dot\alpha}$.

In the R sector, the conjugate of the superghost part $\ee^{-\frac{1}{2}\phi}$, which
is typically present in the vertex, is
$\ee^{-\frac{3}{2}\phi}$ due to the background charge of the chiral boson $\phi$, and
thus we cannot immediately deduce the behaviour of the polarizations
by comparing the conjugated vertices. Nevertheless, the space-time character of the
conjugated polarization is determined, so that (up to a phase) consistent conjugation
rules can be declared.
Our rules are the following (in matrix notation w.r.t. to Chan-Paton indices)
\begin{equation}
\label{conjR}
\begin{aligned}
(\Lambda_{1})^\dagger = \ii \Lambda_{2}~~~&,~~~
(\Lambda_{2})^\dagger = \ii \Lambda_{1}~~,
\\
(M'_{1})^\dagger  = \ii M'_{2}~~~&,~~~
(M'_{2})^\dagger = \ii M'_{1}~~,
\\
\mu^\dagger  = \ii \bar\mu~~~&,~~~
(\bar\mu)^\dagger = \ii \mu~~.
\end{aligned}
\end{equation}
The above relations account for the reality properties of the amplitudes
and solutions appearing in the main text.


\vspace{0.5cm}

\begin{thebibliography}{99}

\bibitem{Seiberg:1999vs}
N.~Seiberg and E.~Witten,
JHEP {\bf 9909} (1999) 032
[arXiv:hep-th/9908142], and references therein.

\bibitem{deBoer:2003dn}
J.~de Boer, P.~A.~Grassi and P.~van Nieuwenhuizen,
Phys.\ Lett.\ B {\bf 574} (2003) 98
[arXiv:hep-th/0302078].

\bibitem{Ooguri:2003qp}
H.~Ooguri and C.~Vafa,
Adv.\ Theor.\ Math.\ Phys.\  {\bf 7} (2003) 53
[arXiv:hep-th/0302109].

\bibitem{Ooguri:2003tt}
H.~Ooguri and C.~Vafa,
``Gravity induced C-deformation,''
[arXiv:hep-th/0303063].

\bibitem{Seiberg:2003yz}
N.~Seiberg,
JHEP {\bf 0306} (2003) 010
[arXiv:hep-th/0305248].

\bibitem{Berkovits:2003kj}
N.~Berkovits and N.~Seiberg,
JHEP {\bf 0307} (2003) 010
[arXiv:hep-th/0306226].

\bibitem{Ferrara:2000mm}
S.~Ferrara and M.~A.~Lledo,
JHEP {\bf 0005} (2000) 008
[arXiv:hep-th/0002084].

\bibitem{Klemm:2001yu}
D.~Klemm, S.~Penati and L.~Tamassia,
Class.\ Quant.\ Grav.\  {\bf 20} (2003) 2905
[arXiv:hep-th/0104190].

\bibitem{Terashima:2003ri}
S.~Terashima and J.~T.~Yee,
JHEP {\bf 0312} (2003) 053
[arXiv:hep-th/0306237].

\bibitem{Britto:2003aj}
R.~Britto, B.~Feng and S.~J.~Rey,
JHEP {\bf 0307} (2003) 067
[arXiv:hep-th/0306215];
R.~Britto, B.~Feng and S.~J.~Rey,
JHEP {\bf 0308} (2003) 001
[arXiv:hep-th/0307091].

\bibitem{Grisaru:2003fd}
M.~T.~Grisaru, S.~Penati and A.~Romagnoni,
JHEP {\bf 0308} (2003) 003
[arXiv:hep-th/0307099];
A.~Romagnoni,
JHEP {\bf 0310} (2003) 016
[arXiv:hep-th/0307209].

\bibitem{Araki:2003se}
T.~Araki, K.~Ito and A.~Ohtsuka,
Phys.\ Lett.\ B {\bf 573} (2003) 209
[arXiv:hep-th/0307076].

\bibitem{Britto:2003kg}
R.~Britto and B.~Feng,
Phys.\ Rev.\ Lett.\  {\bf 91} (2003) 201601
[arXiv:hep-th/0307165].

\bibitem{Lunin:2003bm}
O.~Lunin and S.~J.~Rey,
JHEP {\bf 0309} (2003) 045
[arXiv:hep-th/0307275].

\bibitem{Berenstein:2003sr}
D.~Berenstein and S.~J.~Rey,
Phys.\ Rev.\ D {\bf 68} (2003) 121701
[arXiv:hep-th/0308049].

\bibitem{Alishahiha:2003kg}
M.~Alishahiha, A.~Ghodsi and N.~Sadooghi,
``One-loop perturbative corrections to non(anti)commutativity parameter of N =
1/2 supersymmetric U(N) gauge theory,''
[arXiv:hep-th/0309037].

\bibitem{Imaanpur:2003jj}
A.~Imaanpur,
JHEP {\bf 0309} (2003) 077
[arXiv:hep-th/0308171];
A.~Imaanpur,
JHEP {\bf 0312} (2003) 009
[arXiv:hep-th/0311137].

\bibitem{Grassi:2003qk}
P.~A.~Grassi, R.~Ricci and D.~Robles-Llana,
``Instanton calculations for N = 1/2 super Yang-Mills theory,''
[arXiv:hep-th/0311155].

\bibitem{Britto:2003uv}
R.~Britto, B.~Feng, O.~Lunin and S.~J.~Rey,
``U(N) instantons on N = 1/2 superspace: Exact solution and geometry of moduli
space,''
[arXiv:hep-th/0311275].

\bibitem{Ferrara:2003xk}
S.~Ferrara and E.~Sokatchev,
Phys.\ Lett.\ B {\bf 579} (2004) 226
[arXiv:hep-th/0308021].

\bibitem{Ivanov:2003te}
E.~Ivanov, O.~Lechtenfeld and B.~Zupnik,
``Nilpotent deformations of N = 2 superspace,''
[arXiv:hep-th/0308012].

\bibitem{Billo:2002hm}
M.~Billo, M.~Frau, I.~Pesando, F.~Fucito, A.~Lerda and A.~Liccardo,
JHEP {\bf 0302} (2003) 045
[arXiv:hep-th/0211250].

\bibitem{DiVecchia:1997pr}
P.~Di Vecchia, M.~Frau, I.~Pesando, S.~Sciuto, A.~Lerda and R.~Russo,
Nucl.\ Phys.\ B {\bf 507} (1997) 259
[arXiv:hep-th/9707068];
P.~Di Vecchia, M.~Frau, A.~Lerda and A.~Liccardo,
Nucl.\ Phys.\ B {\bf 565} (2000) 397
[arXiv:hep-th/9906214].

\bibitem{Friedan:1985ge}
D.~Friedan, E.~J.~Martinec and S.~H.~Shenker,
Nucl.\ Phys.\ B {\bf 271} (1986) 93.

\bibitem{DiVecchia:1996uq}
P.~Di Vecchia, L.~Magnea, A.~Lerda, R.~Russo and R.~Marotta,
Nucl.\ Phys.\ B {\bf 469} (1996) 235
[arXiv:hep-th/9601143].

\bibitem{Atiyah:ri}
M.~F.~Atiyah, N.~J.~Hitchin, V.~G.~Drinfeld and Y.~I.~Manin,
Phys.\ Lett.\ A {\bf 65} (1978) 185.

\bibitem{Dorey:2002ik}
N.~Dorey, T.~J.~Hollowood, V.~V.~Khoze and M.~P.~Mattis,
Phys.\ Rept.\  {\bf 371} (2002) 231
[arXiv:hep-th/0206063].

\bibitem{Polchinski:fq}
J.~Polchinski,
Phys.\ Rev.\ D {\bf 50} (1994) 6041
[arXiv:hep-th/9407031].

\bibitem{Green:1997tv}
M.~B.~Green and M.~Gutperle,
Nucl.\ Phys.\ B {\bf 498} (1997) 195
[arXiv:hep-th/9701093];
M.~B.~Green and M.~Gutperle,
JHEP {\bf 0002} (2000) 014
[arXiv:hep-th/0002011].

\bibitem{Frenkel-Kac}
I.~B.~Frenkel and V.~G.~Kac,
Invent.\ Math.\  {\bf 62} (1980) 23.

\bibitem{Kostelecky:1986xg}
V.~A.~Kostelecky, O.~Lechtenfeld, W.~Lerche, S.~Samuel and S.~Watamura,
Nucl.\ Phys.\ B {\bf 288} (1987) 173.

\bibitem{Dixon:jw}
L.~J.~Dixon, J.~A.~Harvey, C.~Vafa and E.~Witten,
Nucl.\ Phys.\ B {\bf 261} (1985) 678;
S.~Hamidi and C.~Vafa,
Nucl.\ Phys.\ B {\bf 279} (1987) 465.

\bibitem{Berkovits:2002zk}
N.~Berkovits,
``ICTP lectures on covariant quantization of the superstring,''
arXiv:hep-th/0209059.

\end{thebibliography}
\end{document}